\documentclass[a4paper, amsfonts, amssymb, amsmath, reprint, showkeys, nofootinbib, twoside,onecolumn,notitlepage]{revtex4-1}

\usepackage[paperwidth=210mm,paperheight=297mm,centering,hmargin=2cm,vmargin=2.5cm]{geometry}

\def\apj{{ApJ}}                 % Astrophysical Journal
\def\apss{{Ap\&SS}}             % Astrophysics and Space Science
\def\aap{{A\&A}}                % Astronomy and Astrophysics
\def\mnras{{MNRAS}}             % Monthly Notices of the RAS
\def\prd{{Phys.~Rev.~D}}        % Physical Review D
\def\nat{{Nature}}              % Nature
\newcommand{\ovp}{\Omega_{vp}}
\newcommand{\oi}{|\dot\Omega_\infty|}
\newcommand{\dt}{\partial_t}
\newcommand{\dx}{\partial_x}

\def\aa{\mathcal{A}_a}

\usepackage{amsmath,amstext}
\usepackage[T1]{fontenc}
\usepackage{amssymb}
\input{epsf}
\usepackage{graphicx}
\usepackage{ae,aecompl}
\usepackage{hyperref}
\usepackage{amsmath}
\usepackage{amssymb}
\usepackage{mathtools}
\usepackage{bm}
\usepackage{cleveref}
\usepackage{tensor}
\usepackage{braket}
\usepackage{enumitem}
\usepackage{mhchem}
\usepackage{color}

\def\be{\begin{equation}}
\def\ee{\end{equation}}
\def\beq{\begin{eqnarray}}
\def\eeq{\end{eqnarray}}

\begin{document}
 
\title{Insights into the physics of neutron star interiors from pulsar glitches}
 
\author{{M.~Antonelli}$^{1}$,~A.~Montoli$^2$,~P.~Pizzochero$^{2,3}$} 

\affiliation{\vspace{2mm}
$^1$~CNRS/IN2P3,~Laboratoire~de~Physique~Corpusculaire,~14050~Caen,~France\\
%$^2$~Nicolaus~Copernicus~Astronomical~Center,~00-716~Warsaw,~Poland\\
$^2$~Dipartimento~di~Fisica,~Universit\`a~degli~Studi~di~Milano,~via~Celora~16,~20133~Milano,~Italy\\
$^{3}$~Istituto~Nazionale~di~Fisica~Nucleare,~via~Celoria~16,~20133~Milano,~Italy
} 

\begin{abstract}
The presence of superfluid phases in the interior of a neutron star affects its dynamics, as neutrons can flow relative to the non-superfluid (normal) components of the star with little or no viscosity. A probe of superfluidity comes from pulsar glitches, sudden jumps in the observed rotational period of radio pulsars. Most models of glitches build on the idea that a superfluid component of the star is decoupled from the spin-down of the normal component, and its sudden recoupling leads to a glitch. This transition in the strength of the hydrodynamic coupling is explained in terms of quantum vortices (long-lived vortices that are naturally present in the neutron superfluid at the microscopic scale). After introducing some basic ideas, we derive (as a pedagogical exercise) the formal scheme shared by many glitch studies. Then, we apply these notions to present some recent advances and discuss how observations can help us to indirectly probe the internal physics of neutron stars. 
\end{abstract} 

\maketitle

\section{Pulsar glitches: general concepts}

This work\footnote{
    Chapter of the volume \href{https://doi.org/10.1142/9789811220944_0007}{\emph{Astrophysics in the XXI Century with Compact Stars}}, eds. C.A.Z.~Vasconcellos and F.~Weber, World Scientific (2022) \cite{AstrophysicsXXIcentury}, submitted in August 2021. The open-access version of this chapter can also be found on the arXiv.org repository: \url{https://doi.org/10.48550/arXiv.2301.12769}.
    Two recent reviews, complementary to this one, are \cite{Zhou_2022Univ,Antonopoulou_review_2022}.
}
is an introduction to the current interpretation and modelling of the glitch phenomenon, a sudden change in the stable rotation period of pulsars. Complementary reviews on the subject are \cite{dalessandro1996review,haskell_review}, while \cite{manchester2018IAUS} gives a contained historical survey of glitch observations since their first detection back in 1969 \citep{Radhakrishnan_first_glitch_vela_1969,Reichley_first_Vela_1969}.
The field is far from being settled, as most of the theoretical studies are still qualitative and based on (Newtonian) phenomenological models. Nonetheless, the glitch phenomenon is probably the clearest observational evidence for the existence of an extended (kilometre-sized) superfluid region in the inner layers of a neutron star (NS). 
Nucleon superfluidity in NSs has found additional support from the real-time monitoring of the rapid cooling of the young NS in Cassiopeia A \cite{Page2011PhRvL,Shternin2011MNRAS}, but this observational result and its interpretation still have to be firmly assessed~\cite{posselt2018ApJ,Wijngaarden2019MNRAS}.

%\vspace{0.3cm}

Pulsar glitches are sudden increases in the observed rotation frequency of a spinning-down neutron star  \cite{lyne2000,espinoza2011,fuentes17,manchester2018IAUS}. 
Soon after the first glitch detection in the Vela pulsar  \cite{Radhakrishnan_first_glitch_vela_1969,Reichley_first_Vela_1969}, \citet{RUDERMAN1969} proposed that the observed spin-up is due to a reduction of the NS's moment of inertia, while
\citet{BAYM1969} associated the observed long post-glitch relaxation timescales -- of order days to months -- with the presence of superfluid neutrons in the interior. 
On even longer timescales, the superfluid may also contribute to another type of rotational irregularity known as timing noise \cite{alpar_noise_1986,jones_noise_1990MNRAS,melatos14noise,antonelli2023MNRAS}, which is commonly observed in pulsars \cite{dalessandro1996review}.
%On even longer timescales, the superfluid may also participate in another kind of rotational irregularity -- known as timing noise \cite{dalessandro1996review} -- that is commonly observed in pulsars \cite{alpar_noise_1986,jones_noise_1990MNRAS,melatos14noise,antonelli2023MNRAS}.

The existence of superfluid phases in NSs was conjectured even before the discovery of pulsars.
Neutron superfluidity is theoretically expected in NS interiors as most of the star will be cold enough for neutrons to undergo spin-singlet Cooper pairing \cite{sauls1989superfluidity,sedrakian2019EPJA}.
Depending on the age of the NS, the interior temperatures are in the range $10^7-10^9\,$K, which are low on the scale defined by the typical Fermi energies of about $10-100\,$MeV (equivalent to $10^{11}-10^{12}\,$K), so that the ratio of temperature to Fermi energy in an NS is comparable to that of laboratory superfluid $^3$He in the millikelvin temperature range.

For what concerns the glitch phenomenon, the presence of a neutron superfluid significantly modifies the internal dynamics of an NS, allowing for the superfluid to flow relative to the ``normal'' (non-superfluid) components. 
For example, the amount of relative flow determines the angular momentum reservoir needed to explain the observed jumps in the pulsar rotation frequency (the glitch amplitude), while the hydrodynamic coupling between the normal and superfluid components modulates the observed slow post-glitch relaxation and, possibly, also the elusive details of the faster spin-up phase.
\\
\\
\indent
The first half of this review revises various aspects of glitch interpretation and modelling.
Then, we move to discuss how to extract information from observations of glitches (Sec. \ref{chap:constraints}, \ref{sec:evolution}). In the last years, several studies have been devoted to demonstrating that the analysis of glitches across the pulsar population has the potential to put constraints on some properties of dense matter. These analyses are tentative first steps, as more high-cadence pulsar timing observations are necessary before it will be possible to falsify some of the theoretical dynamical models (Sec. \ref{sec:stronger}). 
However, glitches provide us with two robust tests for theoretically calculated microscopic inputs (Sec. \ref{sec:max_glitch}, \ref{sec:activity}). In particular, the possibility of inferring physical information from observations of the glitch spin-up phase is presented in Sec. \ref{sec:evolution}.

\subsection{An analogy with type-II superconductors: vortex pinning and creep}
 
To date, the most promising explanation of the glitch mechanism is based on a seminal idea of \citet{ANDERSON1975}, which is built on the analogy between the neutron current that may develop in the inner crust of an NS and the persistent electron current in a type-II superconductor \cite{anderson_kim_64}. 
It is known that a type-II superconductor pierced by a magnetic field can sustain electron currents in a non-dissipative state as long as the vortices carrying a quantum of magnetic flux are anchored to the crystalline structure of the specimen -- i.e. when the quantized flux tubes are ``pinned''. On the contrary, the superconductor enters a resistive state when the flux tubes unpin and start to move \cite{Blatter_review_1994}, creeping through the crystalline structure.

Similarly, in an NS a current of neutrons can flow without resistance as long as the quantized vortex lines, which are naturally present in a rotating superfluid, find energetically favourable to pin to some inhomogeneity in the medium. In this way, it is possible to sustain a persistent current of neutrons that can be dissipated only when the vortices start to creep after some unpinning threshold is reached. According to \cite{ANDERSON1975}, this neutron current is the momentum reservoir needed to explain glitches.

It is believed that vortex pinning can occur in the inner crust and, possibly, in the outer core (with different intensities in different layers). In the non-homogeneous environment of the inner crust, pinning can be with the ions that constitute the crustal lattice and, at least in principle, also with impurities, defects, crystal domain boundaries or pasta structures \cite{alpar77,Donati2004a,link1996,bulgac2013PhRvL,seveso_etal16,klausner_PRC_23}. 

If the protons in the outer core form a type-II superconductor \cite{Baym1969_typeII}, pinning of neutron vortices to the quantized proton flux tubes may be a viable possibility as well \cite{Alpar_pinningCore_2017,sourie2021MNRAS}. 
However, this kind of vortex-flux tube pinning in the outer core is, possibly, even less studied than the already difficult issue of pinning in the inner crust.
The macroscopic behaviour of the proton superconductor may be more complex than that of a typical laboratory superconductor because of the interaction with the neutron superfluid \cite{alford_good_PRB,drummond2017_I,HaberPhysRevD2017,woods2020arXiv}. 
Moreover, the topological excitations threading the spin-triplet neutron superfluid expected in most of the outer core can differ from the vortices of the spin-singlet Cooper pairing case realized in the crust \cite{Brand_3P2,Leinson2020}.  

%Regardless of whether the pinning is realized in the inner crust or the outer core, only when the quantized vortices unpin the superfluid and the rest of the star can interact at the hydrodynamic level via a dissipative force, the so-called \emph{vortex-mediated mutual friction} \cite{Hall1956,Hall1956II,Vinen1957}, which tends to drive the system towards a new metastable state.  

Regardless of whether the pinning occurs in the inner crust or outer core, the superfluid and the rest of the star can interact at the hydrodynamic level via a dissipative force -- known as \emph{vortex-mediated mutual friction} \cite{Hall1956,Hall1956II,Vinen1957} -- only when the quantized vortices unpin and are free to move. Then, mutual friction will drive the system towards a new metastable state.  

To complete the above qualitative picture, it is necessary to provide a reasonable argument for how a current of neutrons should develop in the first place. The starting point is that a rotating superfluid can spin down only by expelling part of its quantized vorticity. However, if the vortex lines are  pinned, their natural outward motion is hindered and the superfluid cannot  follow the spin-down of the rest of the star \cite{ANDERSON1975,pines_alpar_1985Nat}.
In this way, the superfluid component naturally tends to rotate faster than the spinning-down normal component, storing an excess of angular momentum. This reservoir of momentum can then be suddenly released during a glitch when several vortex lines are expelled from the superfluid domain\footnote{
    More precisely, the unpinned vortices will tend to move outward -- away from the rotation axis -- but their displacement is not necessarily uniform across the superfluid region \cite{Alpar1984a,antonelli17,Howitt_Nbody_2020}.
} in a vortex ``avalanche'' involving up to $\sim10^{12}$ out of the $\sim10^{18}$ vortices expected in a fast radio-pulsar. 
Clearly, the same set of ideas also applies in reverse to explain ``anti-glitches'' in spinning-up pulsars, namely sudden spin-up events during which the creep motion of vortices is directed inward~\cite{ducci2015,ray2019ApJ}.

\subsection{Collective vortex dynamics} 

The glitch phenomenon is probably the result of a competition between pinning and the hydrodynamic lift -- the so-called \emph{Magnus force} -- felt by pinned vortices because of the background flow of neutrons \cite{ANDERSON1975}.
Interestingly, vortex avalanches -- the sudden unpinning of several vortices -- occur routinely in two-dimensional simulations of many vortices \cite{Howitt_Nbody_2020}, when chains of unpinning events are triggered collectively in the most unstable parts of the vortex configuration \cite{Warszawski_unpinning}. Similar behaviour is also observed in smaller-scale simulations based on the Gross-Pitaevskii equation \cite{Warszawski_GPE_2011}.
    
\begin{figure}
    \centering
    \includegraphics[width=0.95\textwidth]{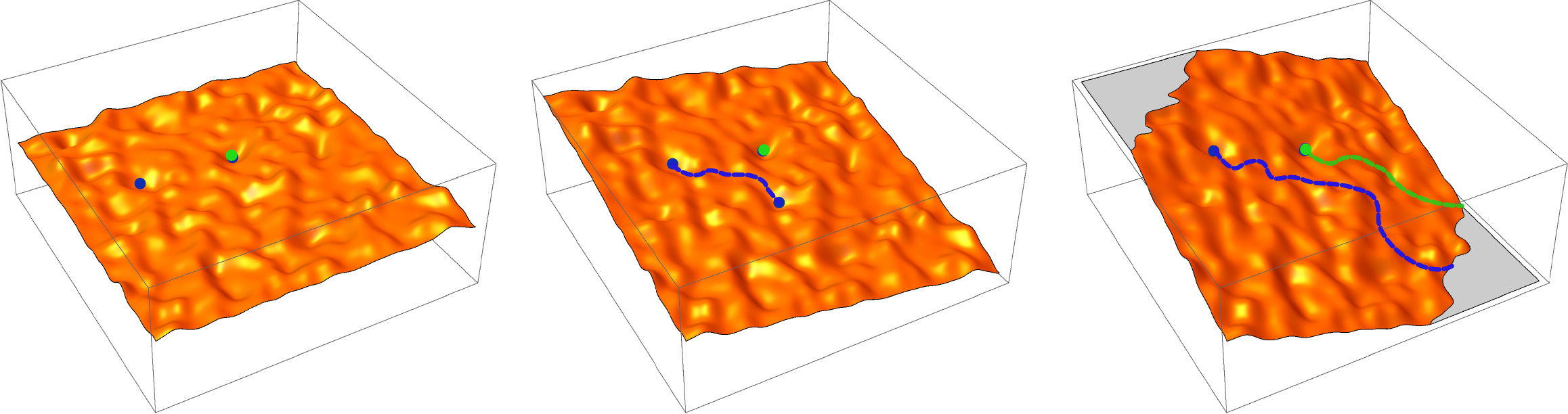}
    \caption{
    Cartoon of the relaxation dynamics between metastable vortex configurations. 
    A particular configuration of vortex lines, here represented by the blue (or green) point, is immersed in the space of all possible configurations. The total energy associated with a configuration defines a disordered landscape over the configuration space. The external torque drives the system and continuously tilts such landscape: when a critical tilt is reached, a metastable configuration can relax to a new minimum (during the relaxation process the tilt angle decreases as the lag between the components is reduced). Since the energy landscape is irregular, different configurations (like the green and the blue points) will start to relax at different tilt angles, following different paths (history dependence). 
    The final configuration depends on how quickly the tilt angle decreases during the process and on the initial configuration itself.
    In principle, the same idea applies in reverse for anti-glitches in spinning-up pulsars.
    }
    \label{fig:pinningladscape}
\end{figure}

This process is reminiscent of the dynamics observed in self-organized critical systems \cite{jensen_book,SOC_2016} and is sketched in Fig.~\ref{fig:pinningladscape}: we can imagine associating a certain energy to each vortex configuration in the inner crust, by accounting for the vortex-vortex and the vortex-lattice interactions. This would, ideally, generate an intricate energy landscape in the space of all possible vortex configurations (reduced to only two dimensions in the figure). 

If no velocity lag between the components is present, then each configuration relaxes to a local minimum by rearranging the mutual positions and shape of the vortices. However, the external torque slowly builds a velocity lag\footnote{
    Following a common practice, we will use the term \emph{lag} to indicate the velocity (or angular velocity) difference between the superfluid and the non-superfluid (i.e. normal) components of the neutron star. In principle, the lag is a local quantity that can vary within the superfluid domain in both space and time.
}, which tends to tilt the energy landscape. Whether this produces a continuous rearrangement of the vortex configuration or a sudden slip to a new metastable state depends on the details of the landscape and of the rearrangement process (i.e. how fast the configuration relaxes with respect to the tilting). 
The idea that glitches are the fingerprint of a self-organized criticality of the vortex configuration \cite{morley1993,MP08} is based on the latter possibility, namely a fast slip to a new, long-lived, metastable state. 

Whether or not glitches would be a genuine manifestation of self-organization in the strict sense \cite{SOC_2016}, \citet{ANDERSON1975} described this driven relaxation process as a noisy stick-slip motion of many vortices, which may be thought of as a sequence of ``quakes'' (slow stress accumulation and fast release) in the vortex configuration.

Despite being just a cartoon, this abstract point of view on the glitch phenomenon allows one to recognize that the sequence of dynamical phases (explored in Sec. \ref{sec:dynamical_phases}) experienced by a pulsar during its slowly driven evolution depends on the properties of an intricate energy landscape. Fig.~\ref{fig:pinningladscape} also suggests that the observed probability distributions of the avalanche sizes (the glitch amplitudes) and waiting times are the manifestations of a stress-release model with multiple thresholds, see e.g. \cite{carlin2021ApJ}. 
In a fast-driven system - i.e., for a larger tilting rate of the landscape in Fig.~\ref{fig:pinningladscape} - it is more difficult for a vortex to settle in a new metastable state. Therefore, is also less likely to develop the intermittent fast-slow dynamics typical of sudden avalanches \cite{jensen_book}, and a smoother and more continuous creep of vortices has to be expected~\cite{Alpar1984a,Baym1992PhyB,antonelli2020MNRAS}.

The nature of the trigger for vortex unpinning is still unclear and proposals range from self-triggered vortex avalanches \cite{ANDERSON1975,pines_1999ptgr,MP08} to hydrodynamic instabilities \cite{Andersson_PRL_twostream,melatos07,Glampedakis2009,Khomenko2019PhRvD} and starquakes -- the failure of the solid crust due to the progressive reduction in the centrifugal force \cite{RUDERMAN1969}, see also \cite{alpar_crab_crack,Franco2000ApJ,giliberti2020MNRAS,Rencoret_quake_2021}. 
During the 2016 Vela glitch, changes in the pulsar's magnetosphere have been observed\footnote{
    Another dramatic change was observed in the high-magnetic field pulsar J1119-6127, where the pulse profile switched from single to double following a large glitch \cite{Weltevrede2011MNRAS}.
}
\cite{palfreyman+2018}, which could be a consequence of the process triggering the glitch, like a crustal quake exciting Alfv\'en waves \cite{bansgrove2020}, or even a clue of the involvement of the magnetic field as a driving force triggering glitches \cite{Ruderman1998}. Some degree of correlation between magnetospheric activity and glitches is also supported by the recent analysis in  \cite{ashton2020arXiv}, which reveals an increased flickering activity of the magnetosphere around the glitch epoch. This result, if confirmed, is interesting, as the flickering activity may be linked to precursors and aftershocks, much like in terrestrial earthquakes. 
If this is the case, the relaxation to a new metastable configuration (see Fig.~\ref{fig:pinningladscape}) could be triggered even before the local critical tilt of the energy landscape is reached, since the quake provides an initial kick -- or ``activation energy''  -- to initiate the avalanche.

Whatever the trigger mechanism, one of the key ingredients in the picture drawn by \citet{ANDERSON1975} is the pinning force that the crustal lattice can exert on a vortex. Similarly to what happens in type-II superconductors, where pinning of flux tubes to impurities in the material enhances the critical current value, the strength of pinning determines the maximum amount of angular momentum that could be exchanged during a glitch \cite{antonelli_GR_2018}. 
The understanding of how much angular momentum can be stored in different regions of the star would allow a comparison with observations of glitching pulsars, potentially constraining the internal structure of an NS \cite{Ho2015,pizzochero17} when used in tandem with other methods based on the study of the average glitch activity \cite{Andersson2012,Chamel2013,carreau+2019,montoli_universe}, dynamical fits of the spin-up phase \cite{graber+2018,ashton+2019,montolimagistrelli+2020} or the glitching history of pulsars \cite{montoli+2020,erbil_vela_2020MNRAS}. Unfortunately, some important details are still missing or are unclear, so not many quantitative conclusions can be drawn to date, as we will discuss in Sec. \ref{chap:constraints} and \ref{sec:evolution}.

\section{Basic formalism for glitch modelling}
\label{sec1.1}

As we have seen, glitch modelling is based on several fruitful analogies between spinning neutron stars and superfluid, or superconducting, systems that are studied in the laboratory \cite{graber_laboratory}.
Moreover, the tools and ideas that are currently used to describe the rotational dynamics of an NS are the byproducts of an attempt to adapt and extend the Tisza-Landau two-fluid model for superfluid $^4$He to account for the several fluid species, and their mutual interactions, present in the interior of an NS \cite{haskell_super,chamel_super,nils_review_2021Univ}.

\subsection{Phenomenological model with  two rigid components }
\label{sec:2c}

The pulsar rotation frequency $\nu(t) = 1/P (t)$ can be measured at a given reference time $t$ using pulsar timing techniques, namely the continuous recording of the time of arrival of each pulse at the telescope.
The precise timing of rotation-powered pulsars reveals a steady and extremely slow increase in the pulse period, indicating that these objects lose angular momentum and kinetic energy due to some emission mechanism.  
However, the rotational frequency of a pulsar occasionally undergoes sudden jumps of amplitude $\Delta \nu > 0$, which are followed by a period of slow recovery that may last for days or months, the so-called glitches. 
The signature of a glitch in the timing residuals is rather clear for relative jumps bigger than $\Delta \nu/\nu \sim 10^{-7}$: when a glitch occurs, the model $P(t)$ for the slow period evolution before the glitch can no longer predict the time of arrivals, which have to be described by a different timing solution by including a discontinuity in  $\nu(t)$ and its derivative $\dot\nu(t)$  around the inferred epoch of the glitch~\cite{lyne2000,espinoza2011}.

Since the NS is an isolated object, the fast spin-up can be explained only in terms of a sudden change of the moment of inertia, or by invoking the action of an internal torque (or both). In particular, the slow post-glitch relaxation of $\nu(t)$ observed in several pulsars is an indication that an internal torque is at work (it is difficult to imagine how a quake changing the moment of inertia could give rise to the post-glitch relaxation). 
In other words,  there is a ``loose'' component inside the star whose rotation is not directly observable, but that is exchanging angular momentum with the observable component.
\\ %some space
\\ %some space
\indent
{\emph{\textbf{A minimal linear model} -} \citet{BAYM1969} proposed the simplest phenomenological description of a glitch: the observable component -- assumed to be rigidly coupled to the magnetosphere, where the pulsar signal originates -- and the loose component in the interior are both modelled as rigid bodies rotating around a common axis. 
They are coupled by an internal torque that is linear in their velocity lag (see \cite{guerci_2017MNRAS,celora2020MNRAS} for nonlinear extensions),
\begin{align}
    \label{baym1}
    & I_o \dot{\Omega}_o + I_s \dot{\Omega}_s = - I |\dot{\Omega}_\infty|
    \\
    & I_s \dot{\Omega}_s = -\frac{I_s \, I_o}{I \, \tau_{\rm{r}}} (\Omega_s-\Omega_o) 
    = -\frac{I_s }{\tau} (\Omega_s-\Omega_o)
        \label{baym2}
\end{align}
where ${\Omega}_s(t)$ and ${\Omega}_o(t) = 2 \pi \nu(t)$ are the angular velocities of the loose component (identified with the superfluid in some, unspecified, internal region) and of the observable component, respectively. The other parameters are the moments of inertia of the two components, $I_s$ and $I_o$, while  $ |\dot{\Omega}_\infty|>0$ sets the intensity of the external spin-down torque, usually linked to electromagnetic radiation. The exact nature of the external torque is unimportant here, but it is assumed to be constant on the timescales of interest. 
The total moment of inertia is $I = I_s+I_o$. 
Since a real NS is not uniform, the quantities that appear in the above equations are interpreted as suitable spatial averages over some regions in the interior, see Sec \ref{circo}. 
Finally, there are two common ways to parametrize the internal torque: in terms of the observed post-glitch exponential timescale $\tau_{\rm{r}}$ (the \emph{relaxation} timescale), or in terms of the \emph{coupling} timescale $\tau$, that is more directly linked to the microscopic processes responsible for the friction between the two components. 

It is interesting to note that the first equation \eqref{baym1} -- representing the total angular momentum balance of the star -- is a fundamental statement, while the modelling enters in the second equation. In fact, equation \eqref{baym2} defines a very simple prescription for the poorly known interaction between the two components (the aforementioned mutual friction), which is responsible for the angular momentum exchange between the two components.
Let us write the initial condition as $\Omega_o(0) = \Omega_o^0$ and $\Omega_s(0) = \Omega_o^0 + \Omega_{so}^0$, where $\Omega_{so}^0$ is the lag at $t=0$. 
Now, the solution of \eqref{baym1} and \eqref{baym2} is
\begin{align}
\label{sol_gen}
    \Omega_o(t) &= \Omega^{\infty}_o(t) + \Delta\Omega_o(t) 
    \\
    \Omega_s(t) &= \Omega^{\infty}_o(t) + \Omega_{so}^0 -  (I_o  /I_s ) \, \Delta \Omega_o(t)  \, .
\end{align}
where $\Omega^{\infty}_o(t) = \Omega_o^0 - t |\dot{\Omega}_\infty|$ describes the steady spin-down, while the angular velocity residue $ \Delta\Omega_o(t) $ reads
\begin{equation}
   \Delta\Omega_o(t)= \frac{I_s}{I} (\Omega_{so}^0-\Omega_{so}^\infty) \big( 1- e^{-t/\tau_{\rm{r}}}\big) \, .
   \label{struzzo}
\end{equation}
Here $\Omega_{so}^\infty = \tau  |\dot{\Omega}_\infty|$ is the constant value of the lag in the steady-state, that is the asymptotic situation in which both components spin down at the same rate $|\dot{\Omega}_\infty|$ imposed by the external torque.  

The general solution  \eqref{sol_gen} may be used to fit the observed post-glitch evolution: assuming that the steady-state spin-down rate $|\dot{\Omega}_\infty|$ and the angular velocity $\Omega_o^0$ are known, we just have to fit the residue in \eqref{struzzo}, an operation that would require the fitting of 2 parameters -- the timescale $\tau_{\rm{r}}$ and the overall amplitude -- that are related to the physical properties of the NS\footnote{
    If the model is used to fit the post-glitch relaxation, then the initial conditions $\Omega_{so}^0$ and $\Omega_o^0$ refer to some time $t=0$ after the unresolved spin-up. From the practical point of view, observations may not be frequent enough to pinpoint the exact instant of the spin-up event. 
}. 

\subsection{Spin-up due to a starquake} 

In their original work, \citet{BAYM1969} assumed that the unresolved spin-up is due to a change in the moments of inertia after the abrupt release of part of the stress accumulated in the crustal solid during the whole NS's life. 
Since the solid crust forms when the frequency of the newborn pulsar is high, it will tend to relax towards a less oblate shape via a sequence of failures (crustquakes) as the star spins down~\cite{RUDERMAN1969}. 
Within this picture, the loose component (the superfluid, whose exact location and extension within the star is unknown), only serves the purpose of explaining why the slow post-glitch relaxation proceeds over the timescale~$\tau_{\rm{r}}$.

We can use the linear friction model in \eqref{baym1} and \eqref{baym2} to study the relaxation after a spin-up induced by a starquake \cite{pines1974,shapiroBook}. 
We already have the general solution \eqref{sol_gen}, so we just have to link the initial condition $\Omega_{so}^0$ to the angular velocity jump following a change in the moments of inertia. The quake can be modelled as an instantaneous  global perturbation at $t=0$, such that
\begin{equation}
\begin{split}
    I_{s,o} \, &\rightarrow \, (1-\epsilon_{s,o}) I_{s,o}
    \\
    \Omega^0_{s,o} \, &\rightarrow \,  \Omega^0_{s,o}/ (1-\epsilon_{s,o})
    \\
    |\dot{\Omega}_\infty| \, &\rightarrow \, (1+\epsilon_{\infty}) |\dot{\Omega}_\infty| \, ,
    \\
    \tau  \, &\rightarrow \, (1+\epsilon_{\tau}) \tau
    \end{split}
    \label{epsilons}
\end{equation}
where $\epsilon_{s,o}>0$ since the star should evolve towards a less oblate shape, while the signs of $\epsilon_{\tau,\infty}$ are undetermined. 
Assuming the star to be in the steady-state before the starquake, the post-glitch relaxation can be written in terms of the pre-glitch parameters and the relative variations $\epsilon$: 
\begin{multline}
        \Omega_o(t) = (1+\epsilon_o)  \Omega_o^0 - (1+\epsilon_{\infty}) |\dot{\Omega}_\infty| t  +
   \frac{I_s}{I} \big[  (\epsilon_s-\epsilon_o) \Omega_{o}^0 -  (\epsilon_\infty+\epsilon_\tau-\epsilon_s)  
         \Omega_{so}^\infty 
         \big] \big( 1- e^{-t/\tau_{\rm{r}}}\big)
    + O(\epsilon^2)
    \label{quillo}
\end{multline}
    where, again, $\Omega_{so}^\infty=|\dot{\Omega}_\infty| \tau$ is the pre-starquake steady-state lag (all the quantities but $\epsilon$ in \eqref{quillo} refer to some instant before the starquake, including  $\Omega_o^0$).

The observed glitch jump $\Delta\nu$ allows us to fix the parameter $\epsilon_o$ as $\epsilon_o = \Delta \nu /\nu$, but it is difficult to estimate the other $\epsilon$  parameters. However, it is possible to relate them to the so-called healing parameter $Q$, which may be obtained from observations: to the first order in $\epsilon$, equation \eqref{quillo} may be expressed as 
\begin{equation}
        \Omega_o(t) =   \Omega_o^0 + \Delta\Omega_o - (1+\epsilon_{\infty}) |\dot{\Omega}_\infty| t  
        -   Q \Delta\Omega_o \big( 1- e^{-t/\tau_{\rm{r}}}\big)
    \label{quilloQ}
\end{equation}
where $\Delta\Omega_o \approx \epsilon_o \Omega_o^0$ is the observed glitch amplitude. Hence, 
\begin{equation}
    Q = \frac{I_s}{I} \left( 
    1-\frac{\epsilon_s}{\epsilon_o} + \frac{ \epsilon_\infty+\epsilon_\tau-\epsilon_s }{ \epsilon_o } 
    \, \frac{ \Omega_{so}^\infty   }{  \Omega_{o}^0  }
    \right) \approx \frac{I_s}{I} \left( 1-\frac{\epsilon_s}{\epsilon_o}\right) \, > \, \frac{I_s}{I} 
    \label{QQQ}
\end{equation}
where the approximated expression is obtained in the limit $\Omega_{so}^\infty  \ll \Omega_{o}^0 $. 
\citet{crawford03} collected all the measured values of the healing parameter $Q$
for Crab and Vela pulsars available in the literature at that time. They found that a weighted mean of the $\sim 20$ measured values of $Q$ for Crab glitches the values yields $Q \approx 0.7 \pm 0.05$.
On the other hand, a mean value for the glitches in the Vela pulsar yields a smaller value, $Q \approx 0.12 \pm 0.07$. 
To test the starquake model for the Crab and the Vela pulsars, they also estimated the healing parameter as $Q \approx I_s/I$ -- which is valid in the limit of $I_o \ll I_s $ and/or $\epsilon_s I_s \ll  \epsilon_o I_o$ -- for different equations of state dense nuclear matter covering a range of neutron star masses, and interpreting $I_s$ as the moment of inertia of the neutrons in the core. 
They conclude that the results are consistent with the starquake model predictions for the Crab pulsar, but the much smaller values of $Q < 0.2$ of the Vela pulsar are inconsistent with the constraint in \eqref{QQQ} since the implied Vela mass would be unreasonably small, about $\sim 0.5 \,  M_\odot$.

\subsection{Some insights from seismology } 

While starquakes may be a plausible explanation for at least the smallest glitches observed in the Crab pulsar, the original idea of \citet{RUDERMAN1969} is challenged by the large glitch activity observed in the Vela \cite{chamelReview,haskell_review,Rencoret_quake_2021}.
Apart from the large amplitudes of the Vela glitches, which would require an unrealistic change in the moments of inertia of the order of $\delta \nu/\nu \approx \epsilon_o \sim 10^{-6}$, the problem is also their frequency (roughly one in $2-3$ years): the spin-down seems to be too slow for the crust to break because of the change in the centrifugal force between two glitches, at least if the crust relaxes to an unstressed configuration \citep{giliberti2020MNRAS}. 
Therefore, starquakes could still be a viable trigger for glitches in the Vela (see e.g., \cite{akbal2017}), but only if~\cite{Franco2000ApJ,giliberti2020MNRAS,Rencoret_quake_2021}:
\begin{enumerate}
    \item[-] the crust never relieves all the accumulated stresses (which is also the case of earthquakes, see Fig.~\ref{fig:quake}),
    \item[-] and/or some internal loading force (possibly due to vortex pinning \cite{Ruderman1976} or to the magnetic field  \cite{Ruderman1998,Franco2000ApJ}) that is orders of magnitude stronger than the variation in the centrifugal force during the inter-glitch time is operating in the crust.
\end{enumerate}
Crustal motion after failure is also invoked as a plausible mechanism to explain magnetar emission. However, it is possible that neutron star matter cannot exhibit brittle fracture at any temperature or magnetic field strength \cite{jones2003_fault}: the crust could deform in a gradual plastic manner despite its extreme strength, leading to rarer starquakes than calculated for a rigid crust \cite{Smoluchowski_plastic_crust}. On the other hand, the presence of defects and crystal domains  \cite{Ruderman1991_Tectonics_I} and the fact that stresses are applied continuously for a long time  \cite{Chugunov_2010MNRAS} are all elements that point in the opposite direction (they decrease the breaking strain), namely toward the possibility of more frequent crust failures. The situation is far from being settled and the very question of whether or not crustquakes actually can happen -- and with what frequency -- in radio pulsars is uncertain (see \cite{Rencoret_quake_2021} for a critical revision of the starquake paradigm). 

However, the analogy between quakes and glitches goes beyond the questions about the nature of the trigger and the magnitude of the change in the moment of inertia in \eqref{epsilons}: there is also a more philosophical link between glitches and earthquakes. 

Both glitches and earthquakes are the byproducts of a slowly driven system that sporadically releases internal stress in a sequence of fast bursts. This is the kind of dynamics expected to emerge in many complex systems, that exhibit rare and sudden transitions occurring over intervals that are short compared with the timescales of their prior evolution \cite{jensen_book}. 
Such extreme events are expressions of the underlying forces\footnote{
    Like the competing pinning force and the Magnus force (the hydrodynamic lift felt by a pinned vortex, proportional to the velocity lag between the normal component and the superfluid one). In this sense, the velocity lag in a pulsar provides a measure of its internal stresses.
}, or stresses, usually hidden by an almost perfect balance.
An earthquake is triggered whenever a fracture (the sudden slip of a fault) occurs in the Earth's crust, as a result of mechanical instability. Likewise, a glitch is triggered by the very same mechanism -- in the case, the trigger is a crustquake -- or by a mechanical instability in the vortex configuration -- i.e., a self-triggered vortex avalanche if we stick to the paradigm of \citet{ANDERSON1975}. 
\begin{figure}
    \centering
    \includegraphics[width=0.9\textwidth]{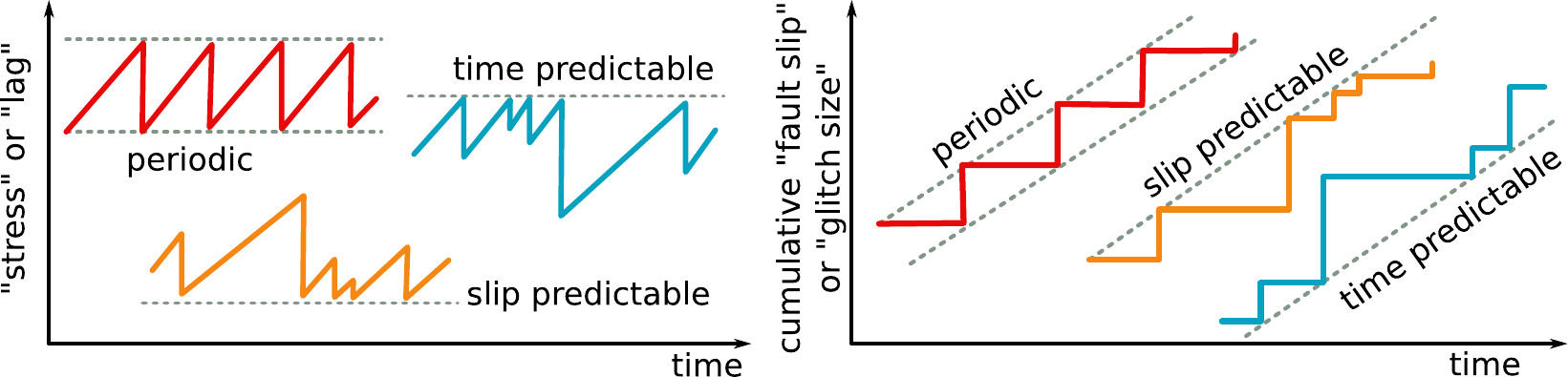}
    \caption{
    The three scenarios for the evolution of tectonic stresses: \emph{periodic} (where the initial $\sigma^c$ and final $\sigma^{low}$ stresses of faulting are constant), \emph{time-predictable} (only $\sigma^c$  is constant, so the larger a quake the longer the following quiet period), \emph{slip-predictable} (the final stress $\sigma^{low}$ is constant, so the longer the quiet period the larger the quake that follows). 
    The same scheme can be applied to the stress accumulated in the crust of an NS and/or in the vortex configuration because of the competition between the Magnus and pinning forces (the velocity lag between the normal and superfluid components is a proxy for the stresses in the vortex configuration). 
    The right panel shows the cumulative observed slip (i.e., the cumulative quake size, or the cumulative glitch amplitude) as a function of time for the three scenarios.  
    }
    \label{fig:quake}
\end{figure}
\\
\\
 \indent
 \emph{\textbf{Glitch forecast: time and slip predictability} - }
Despite there is an opinion that earthquakes -- and so glitches -- could be inherently unpredictable, observations have shown that the process of seismogenesis is not completely random.
The fact that earthquakes seem to be clustered in time more than would be expected for a random process (Omori's law describes the increased rate of small earthquakes after large ones) is considered to be expression of a degree of determinism and predictability in the properties of the earthquake population \cite{Sornette_2002PNAS}. 
In particular, \citet{Shimazaki1980Geo} identified the three scenarios shown in Fig.~\ref{fig:quake}:
\begin{enumerate}
    \item[] \emph{Time-predictable -} Assume that the stress $\sigma(t)$ along a fault builds up at a nearly constant rate $\dot{\sigma}$ and that a certain amount of stress -- resulting in a slip of the fault -- is released when a certain (constant in time) critical threshold $\sigma^{c}$ is topped. If we observe a drop in the stress $\Delta\sigma$, then the time to the next quake will be $\Delta t \sim \Delta\sigma/\dot{\sigma}$.
    Even if $\dot{\sigma}$ is not directly observable, it can be inferred by observation of the previous waiting times between quakes and their amplitudes so that we can estimate $\Delta t$.
    \\
    \item[] \emph{Slip-predictable -} The critical threshold $\sigma^{c}$ may not be constant in time: the fault itself  -- or the vortex configuration -- is spatially non-homogeneous and it may also rearrange in time (because of small rearrangements during the continuous stress build-up or modifications induced by past quakes). 
    However, if the stress in a quake is released till it drops below a certain level $\sigma_{low}$ that is fairly constant in time, then it is possible to predict the size of the quake: 
    $\Delta\sigma \sim \dot{\sigma} \, \Delta t $, where now $ \Delta t$ is the observed time elapsed since the previous quake.
    \\
    \item[] \emph{Periodic -} When both thresholds  $\sigma^{c}$ and  $\sigma^{low}$ remain almost constant in time, the stress release works as a clock: the distributions of both waiting times and sizes should be peaked around a well-defined value.
\end{enumerate}
Observations of a sequence of events may reveal a tendency of the system for one of these three scenarios -- or for the most general one, in which neither $\sigma^{c}$ or $\sigma^{low}$ is constant. This is shown in the right panel of Fig.~\ref{fig:quake}: information is carried by the shape of the cumulative activity constructed from the observed sequence of quakes, or glitches.

Glitch forecast may not be as socially important as earthquake forecast but observing glitches requires constant monitoring effort. Hence, attempting to provide simple estimates of the time between two successive glitches might also have practical value in optimizing observations, as well as possible falsification of the statistical model used \cite{Itoh_1983_predictions,middleditch2006,akbal2017,Melatos2018ApJ,montoli+2020}. 
%Glitch forecast may not be as socially important as earthquake forecast, but, given the fact that observing glitches requires a constant monitoring effort, providing at least rough estimates of the time till the next glitch could also have some practical value \cite{Itoh_1983_predictions,middleditch2006,akbal2017,Melatos2018ApJ,montoli+2020}. 
Unfortunately, to date, a well-defined statistical procedure for predicting a glitch in a given pulsar has not yet been developed: we still have a lot to learn about the statistics of glitch occurrence in single objects. 
\\
\\
\indent
\emph{\textbf{Glitch statistics, self-organization and hysteresis} -}
While statistical analysis of the whole glitch population has been carried out by aggregating data from different pulsars \cite{sherman1996,lyne2000,fuentes17}, studies of the glitch statistics for individual pulsars are considerably more difficult because of small number statistics. Nonetheless, as the number of detected glitches increases, it is becoming possible to have a rough idea of the probability density functions for the glitch sizes and the waiting times in a given pulsar \cite{ashton17,howitt2018ApJ,fuentes2019}. Most pulsars are roughly compatible with exponentially distributed waiting times and power-law distributed event sizes, which is the expected behaviour arising from self-organized critical systems\footnote{
    Self-organized criticality is believed to provide a natural explanation of the statistics of earthquakes, including the Gutenberg-Richter law for the distribution of earthquake magnitudes \cite{Bak_earthquakes_1989,SornetteSornette1989,Ito_SOCquake_1990,bak1994GMS}.
} \cite{MP08}. However,  the statistical properties of the Vela pulsar and J0537-6910 are unique among glitching pulsars, as they typically have large glitches which occur fairly periodically. Their observed activity is shown in Fig.~\ref{fig:fit_activity}.

\begin{figure}
    \centering
    \includegraphics[width=0.8\textwidth]{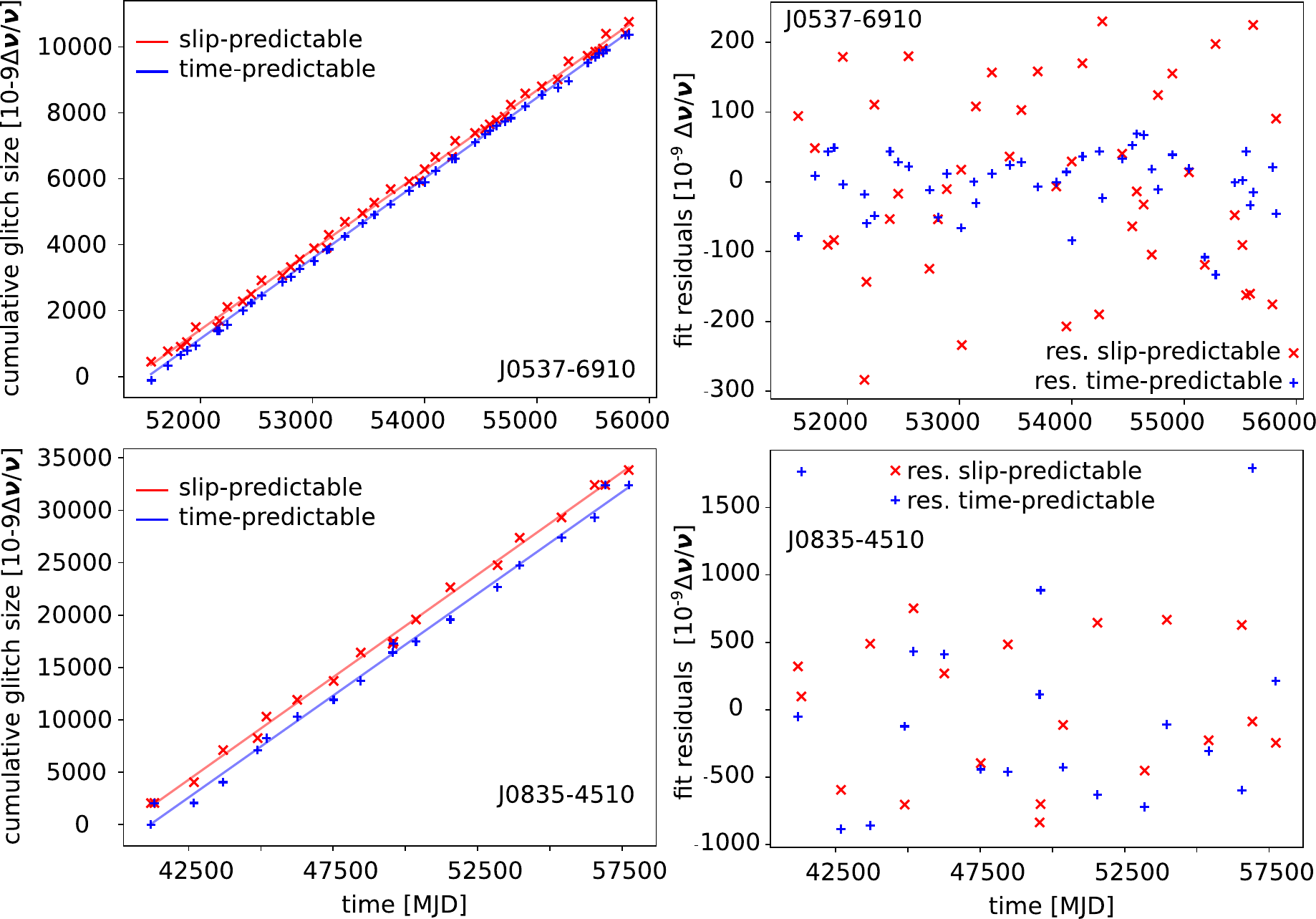}
    \caption{
    The glitch activity of J0537-6910 and the Vela pulsar (J0835-4510). The upper and lower points of the cumulative glitch amplitude have been fitted with the standard linear regression procedure to test whether those two quasi-periodic pulsars display a tendency for the slip-predictable or time-predictable scenarios. The residuals of the fits are shown in the right panels. For J0537-6910 the residuals relative to the slip-predictable fit line are more dispersed than the ones relative to the time-predictable case, indicating some slight preference of the system for the latter. 
    }
    \label{fig:fit_activity}
\end{figure}

In particular, J0537-6910 is one of the most actively glitching pulsars we know of, probably because of its high spin-down rate (see \cite{Ho_big_glitcher_2020} and \cite{Akbal2021arXiv} for a recent summary of its properties). 
Its glitching behaviour seems to be particularly favourable to glitch prediction due to the presence of some degree of correlation between the waiting time from one glitch to the next and the amplitude of the first glitch \cite{middleditch2006,Ferdman2018ApJ,Antonopoulou2018MNRAS}. 
The analysis of its cumulative glitch amplitude in Fig.~\ref{fig:fit_activity} also reveals an imprint of such a time-predictable behaviour. 

Apart from the case of J0537-6910 (and, to a lesser extent, J1801-2304), only a few statistically compelling correlations are found between the size of a glitch and the time to the preceding or the succeeding glitch \cite{Melatos2018ApJ}. Given that glitches are likely to be threshold-triggered events, this may sound counterintuitive.  
However, this absence of obvious correlations can be reproduced in terms of a state-dependent Poisson process for the internal stress $\sigma(t)$, where the rate with which a glitch is triggered diverges as the internal stress reaches a threshold $\sigma_c$ \cite{Fulgenzi2017,Carlin2019MNRAS}. Alternatively, Fig.~\ref{fig:quake} suggests that the lack of obvious correlations may be due to a process that conserves neither $\sigma^c$ nor $\sigma^{low}$. 

The non-constancy of the upper (unpinning) and lower (repinning) thresholds is naturally explained within the vortex avalanche scenario. In fact, the vortex configurations are expected to move from a local minimum to another in a complex landscape, as sketched in Fig.~\ref{fig:pinningladscape}: each starting minimum would be associated with a different  $\sigma^c$, and every final position in the pinning landscape to a certain $\sigma^{low}$. Unsurprisingly, the exact sequence of such values would depend on the past history of the system -- also in view of the possible presence of \emph{hysteresis}\footnote{
    Note added to the arxiv version: hysteresis and properties of the pinning landscape are also explored in \cite{link2022ApJ}, thanks to simulations of vortex filaments similar to those developed in~\cite{link2009}. 
} in the vortex configuration \cite{antonelli2020MNRAS} -- and only very long timing observations may unveil some of its statistical properties. 
As the number of glitches detected in single pulsars grows enough to allow systematic and reliable studies of the statistical properties of the glitching behaviour, it will be possible to study how these properties change across the pulsar population (i.e. how glitch properties depend on the estimated pulsar age, magnetic field, temperature and spin-down rate).

\section{Two-fluid equations and mutual friction}
\label{sec:fluid}
 
In order to incorporate the seminal ideas of \cite{BAYM1969,ANDERSON1975} in more refined models, much attention has been devoted to developing fluid theories for the macroscopic degrees of freedom of neutron star matter. This is necessary if we want to go beyond the minimal rigid-body model of  Sec.~\ref{sec:2c}.

In the presence of superfluid phases, the set of thermodynamic variables needed to define a homogeneous equilibrium state must be enlarged to account for the possible presence of long-lived metastable states with persistent currents \citep{gavassino2020CQG}. It follows that the system is a \emph{multifluid}, a conducting medium where different chemical species (or even abstract species, like the entropy) can flow with respect to each other~\cite{carter1989}. 
The multifluid formalism necessary to build glitch models is based on the seminal works of Carter and collaborators -- e.g., \cite{carter92,langlois_etal98} -- and has been recently reviewed in \cite{haskell_super,chamel_super,nils_review_2021Univ}. 
An extended presentation of relativistic multifluids can be found in \cite{AnderssonLivRev}, while a dictionary to translate between different multifluid theories is given in \cite{gavassino2020CQG}. 

Superfluidity and superconductivity in NSs and nuclear matter, with a focus on pairing phenomena, are reviewed in \cite{sauls1989superfluidity,sedrakian2019EPJA}: from a practical point of view, we only need to recall the fact that neutrons in the inner crust can be described at the hydrodynamic level by a scalar complex order parameter \cite{chamel_super}.  
In fact, in the inner crust, neutrons are expected to form Cooper pairs in the $^1S_0$ channel of the neutron-neutron interaction potential, implying that the superfluid order parameter in the crust is a scalar. Therefore, the superfluid in the crust is expected to behave -- at least in the homogeneous limit, where the crustal nuclei are neglected -- similarly to superfluid $^4$He, whose hydrodynamics is well described by the so-called HVBK hydrodynamics~\cite{Qbook,soninBook16,GusakovHVBK,GAVASSINO_iordanskii}. 

For what concerns the superfluid neutrons in the core, according to the current understanding of nuclear interactions obtained from nucleon scattering data, pairing is more likely to occur in the $^3P_2$ channel above nuclear saturation \cite{Fujita_1972,Yasui_2019PhRvC}, so that the order parameter is anisotropic~\cite{Brand_3P2}.
In the following, we will neglect the issues that arise when the order parameter for neutron superfluidity is not scalar, like the possible presence of domain walls \cite{Yasui_PRC2020},   vortices of half-integer circulation and spin textures \cite{Leinson2020}. 
Despite these complications, some hydrodynamic models for the neutron superfluid and proton superconductor mixture in the outer core have also been proposed, e.g.~\cite{mendell1991ApJ_I,mendell1991ApJ_II,mendell1998,glampedakis11,Gusakov2016,Rau_wass_2020,sourie2021MNRAS}: a key issue is how to effectively account for the presence of neutron vortices interacting with many proton flux tubes, at least if we are in a part of the outer core where the $^1S_0$ pairing may be realized \cite{Alpar_pinningCore_2017,drummond2017_I}.
\\
\\
\indent
\emph{\textbf{Fluid components} -} 
Almost all glitch models to date are based on the assumption that (locally) it is possible to model neutron conduction in terms of two components: a first fluid that should represent the neutron superfluid (labelled by the index $x=n$) and an effective charge-neutral fluid ($x=p$). 
This second component is normal (not superfluid), in the sense that, when the system is in thermodynamic equilibrium, its comoving reference frame is the one in which the entropy of the multifluid is defined~\cite{gavassino2020CQG}. This normal component is a neutral mixture of charged particles and, possibly, some neutrons that are strongly coupled to the protons (for example, the ones that we assume to be bound in the lattice of the inner crust \citep{carter_macro_2006}). All these species can be combined into the p-component as long as charge neutrality is satisfied over macroscopic regions (local charge imbalances are expected to be equilibrated by the electron fluid on very short timescales).
In the following, we adopt this simplified zero-temperature effective description, where the local generation of entropy by friction and the resulting heat diffusion is not taken into account\footnote{
    Adding heat would require a 3-current model (superfluid, normal and entropy currents, see \cite{Gavassino2022_3current} and references therein):  
    the entropy current is locked to the typical p-component only when there is no heat flux, which reduces the 3-current model to a 2-current one. % equation (94) Gavassino2022_3current
    However, heat diffusion is important to study the possibility of having thermally-driven glitches, where the heat generated by vortex motion affects the vortex creep rate~\cite{link1996,Larson2002MNRAS}.
}.
\\
\\
\indent
\emph{\textbf{Fluid equations in the non-transfusive limit} -}  We adopt the Newtonian multifluid formalism of \citet{P04}, in the limit where there are no chemical reactions involving the two species (see  \cite{chamel_carter_2006MNRAS,gusakov2006MNRAS,andersson2011IJMPD} for equivalent formulations and \cite{gavassino2020CQG} for a dictionary to map one into the other). 
This simplified description is the most commonly used setting when it comes to glitch modelling. 
For  $x =n,p$, the hydrodynamic equations are:
\begin{align}
&	\partial_t \rho_x + \nabla_i  ( \rho_x v^x_i ) = 0  
	\label{eq:continuity}
\\
&	\rho_x \left( \partial_t + L_{v_x} \right) p^x_i
	+ \rho_x \nabla_i \left( 
	%\Phi + 
	\tilde{\mu}_x - |v_{x}|^2/2 \right) = F_{i}^{x} \, ,
	\label{eulerox}
\end{align}
where $L_{v_x}p_i^x$ is the  Lie derivative of the momentum per particle $p_i^x$ along the field $v_i^x$,
\begin{equation}
	L_{v_x} p_i^x = v_x^k \partial_k  p_i^x +  p_j^x \partial_i v_x^j \, .
\end{equation}
In the momentum equation \eqref{eulerox},  $\tilde{\mu}_x$ is the specific chemical potential of each component and $\rho_x=m_x n_x$ is the mass density, where $m_x$ is the baryon mass and $n_x$ is the associated baryon number density. The effect of gravity or external fields can be included in  the force density $F_{i}^{x}$. The arbitrariness related to how to count the baryons in $n_x$ is referred to as ``chemical gauge'', and is discussed in, e.g.,  \cite{carter_macro_2006,GAVASSINO_iordanskii} and references therein. 
%For simplicity, we will make the approximation $m_p = m_n$.
\\
\\
\indent
\emph{\textbf{Velocity, momentum, entrainment} -} The field $v_x^i$ is sometimes called kinematic velocity to stress the fact that it is the quantity that appears in the conservation equations \eqref{eq:continuity}. On the other hand, the momentum of the superfluid species $p_i^n$ is, at the inter-vortex scale, proportional to the gradient of the phase of the superfluid order parameter, so that it is an irrotational field at the inter-vortex scale~\cite{carter92,langlois_etal98}. 
Hence, $p_i^n$  is analogous to the ``superfluid velocity'' of the Tisza-Landau or HVBK\footnote{
    In the HVBK extension of the Tisza-Landau two-fluid model, the ``superfluid velocity'' is  coarse-grained over many vortices \cite{Qbook}. Analogously, at the macroscopic scale, $p_i^n$ is the averaged momentum in a fluid element containing several vortex lines. 
} models for superfluid $^4$He~\cite{andersson2011IJMPD,GusakovHVBK,gavassino2020CQG}. 

To gain some intuition on the distinction between velocity and momentum we just have to recall their role. The velocity field $v_x^i$ can be defined starting from the conserved (in the non-transfusive limit considered here) current: this field, or better, its associated current, is related to the counting of particles carrying the certain label $x=n,p$. 
The momentum, on the other hand, defines the system's response to a force. 
Assume a homogeneous system where there are no gradients and for which we have already defined the velocities $\mathbf{v}_n$ and  $\mathbf{v}_p$ via a chosen counting procedure. 
We can now make a thought experiment and apply an external force $\mathbf{F}^{ext}_n$ only to, say, the n-component: clearly, we expect $\mathbf{v}_n$ to change, but, thanks to the interaction between the two species, also $\mathbf{v}_p$  may change by some amount. Hence, it is reasonable to write
\begin{equation}
    \mathbf{F}^{ext}_n =n_n \dot{\mathbf{p}}_n = \rho^*_{nn} \dot{\mathbf{v}}_n+ \rho^*_{pn} \dot{\mathbf{v}}_p \, ,
    \label{pincopallo}
\end{equation}
where $\rho^*_{nn}$ and $ \rho^*_{pp}$ depend on the details of the microscopic interaction.
In general, for a force that is applied only to the $x$-component, we have $ \mathbf{F}^{ext}_x =n_x \dot{\mathbf{p}}_x = \rho^*_{nx} \dot{\mathbf{v}}_n+ \rho^*_{px} \dot{\mathbf{v}}_p $: the four effective densities $\rho^*_{xy}$ for $(x,y=n,p)$ are response functions\footnote{
    In general, the $\rho^*_{xy}$, which are not all independent, are nonlinear functions of the velocities~\cite{Leinson2017MNRAS}. They were first introduced to model $^4$He-$^3$He superfluid mixtures~\cite{khala57,andreevbashkin1976}.
}, which may be measured by keeping track of the acceleration of a component under an applied external force. 
As we will see, this is exactly what happens in glitch models, where we have an external force (the spin-down torque) that only acts on the p-component and we are interested in the response of both components.

The last piece of the puzzle that is missing is the following: why does it happen that $\mathbf{p}_n$, and not $\mathbf{v}_n$, is the quantity that is related to the gradient of the phase of the order parameter?  This is a self-consistency requirement for the hydrodynamic theory at the inter-vortex scale \cite{carter92}, see Sec. 3 of \cite{gavassino2020CQG}. 
The intuition behind this fact was first discussed by \citet{anderson1966} on the basis of the Josephson effect for superfluid $^4$He.

In the end, the relation \eqref{pincopallo} between $p_i^x$ and $v^i_x$ turns out to be a convex combination of the two velocity fields, that can be written in terms of two dimensionless parameters $\epsilon_x$~\cite{P04},
\begin{equation}
\label{entr_pn}
    p_x^i = m_x \left[ (1-\epsilon_x) v_x^i + \epsilon_x  v_y^i \right]
    \quad 
    \text{for} \quad x \neq y \, ,
\end{equation}
where $\rho_n \epsilon_n = \rho_p \epsilon_p$ to guarantee some fundamental properties of the hydrodynamic stress tensor $T^{ik}$ to be discussed below~\cite{carter_macro_2006}.

The leading microscopic processes that are responsible for the exact value of $\epsilon_x$, or $\rho^*_{xx}$, depend on the local properties of matter within the star.
In NS cores entrainment is due to the nuclear interaction, that dresses a proton with a cloud of neutrons \cite{Gusakov2005NuPhA,chamelhaensel2006,Leinson2017MNRAS,chamel2019PRC}. 
In the inner crust, entrainment arises from the interaction of the free neutron gas with the nuclear clusters in the lattice or pasta phases and, in close analogy with electrons in a metal, band-structure effects~\cite{CC05,chamel2012}. 
\\
\\
\indent
\emph{\textbf{Chemical gauge} -} The explicit presence of the entrainment parameters makes clear that the momenta $\mathbf{p}_x$ and the kinematic velocities $\mathbf{v}_x$  are not the same thing. 
These quantities may depend on the prescription used to divide the total baryon density into $n_x$, the densities associated with the two effective components. This arbitrariness is sometimes referred to as chemical gauge choice.

While in the core we have the natural chemical gauge choice of considering $\rho_n$ to be the total mass density of all neutrons, this is not always the case in the inner crust, where different gauge choices can be made. 
In some works, $\rho_n$ refers to the density of the dripped neutrons, the energetically unbound ones (the ``conduction'', or ``free'' neutrons). 
The values of the entrainment parameters  $\epsilon_x$ depend on this gauge choice. 
In particular, $\epsilon_n$ defines the mobility of the n-component and, ultimately, the effective density 
\begin{equation}
    \rho^*_n = \frac{\rho_n}{1-\epsilon_n}
    \label{marzullo}
\end{equation}
that has to be ascribed to the part of neutrons that are effectively free to move \cite{carter_macro_2006}. 
The density $\rho^*_n$ is analogous to the phenomenological ``superfluid density'' that appears in the standard formulation of the Tisza-Landau or HVBK models of superfluid $^4$He~\cite{P04,carter_macro_2006,chamel_carter_2006MNRAS,andersson2011IJMPD,GAVASSINO_iordanskii}.
Interestingly, it is possible to show that $\rho^*_n$ does not depend on the particular chemical gauge choice \cite{carter_macro_2006,GAVASSINO_iordanskii}.
In other words, $\rho^*_n$ does not depend on the ambiguity regarding how many neutrons should be counted as free in the inner crust. For this reason, we will use $\rho^*_n$ to define the effective moment of inertia of the superfluid in glitch models \cite{chamel_carter_2006MNRAS,antonelli17}.
\\
\\
\indent
\emph{\textbf{Mutual friction} -} In glitch models, the force $F_{i}^{x}$ in \eqref{eulerox} is the so-called \emph{mutual friction}, which allows for momentum transfer between the two components.
This force was first studied and proposed to model the dynamics of superfluid $^4$He  \cite{Hall1956,Hall1956II,Vinen1957}. The adjective ``mutual'' emphasises the fact that, in the absence of external forces (i.e. $F_{i}^{n}+F_{i}^{p}=0$), the total momentum density 
\begin{equation}
j_i = n_n p_i^n + n_p p_i^p = \rho_n v_i^n + \rho_p v_i^p     
\label{jtot}
\end{equation}
is conserved, in the sense that 
\begin{equation}
\partial_t j^i + \nabla_k T^{ik}  =0 \qquad T^{ik} = \sum_x  n_x v_x^i p_x^k + \Psi \delta^{ik} 
\, ,
\label{prezzemolo}
\end{equation}
where $\Psi$ is a thermodynamic potential that reduces to the usual pressure when the two species comove. The symmetry $T^{ik}=T^{ki}$ is guaranteed by the before-mentioned property $\rho_n \epsilon_n = \rho_p \epsilon_p$.

First, we have to understand why the mutual friction $F_{i}^{x}$ can be related to the presence of vortices. The idea is simple: for any given vortex configuration, the phase of the order parameter is assigned. If the vortex configuration is frozen, the phase field can not change, nor its gradient. It follows that, in order to change $\mathbf{p}_n$, we have to displace some vortices\footnote{
    At the mesoscopic scale, the momentum $\mathbf{p}_n$ is proportional to the gradient of the order parameter's phase \cite{carter92,gavassino2020CQG}.
    }. 
Hence, it is reasonable to expect that $\mathbf{F}^{n}$, which exactly defines $\dot{\mathbf{p}}_n$ in the homogeneous limit, can be written as a function of the local averaged velocity of vortices and their number density \cite{antonelli2020MNRAS,andersson_MF}. 
In particular, we can expect $\dot{\mathbf{p}}_n$, and so $\mathbf{F}^{n}$, to be proportional to the average ``vortex current'' in the fluid element. 
The problem is how to define the concept of current for string-like objects, but it is possible to do so in the usual way if the vortices are \emph{locally} parallel.
Under this hypothesis, the macroscopic vorticity 
\begin{equation}
   \boldsymbol{\omega} = \nabla \times \mathbf{p}_n /m_n = \kappa n_v \hat{\boldsymbol{\omega}}
   \qquad 
   \kappa n_v = |\boldsymbol{\omega}|
   \label{asdrubale}
\end{equation}
reflects closely the vortex arrangement in the local fluid element. In the above equation, $\kappa$ is the quantized circulation of the momentum around a single vortex, $n_v$ is the areal density of vortices measured in a plane orthogonal to $\hat{\boldsymbol{\omega}}$, the local direction of the vortices.
In other words, the hypothesis that the quantum vortices are locally parallel is so strong that there is almost no loss of information in performing the average procedure over the fluid element to obtain the macroscopic momentum $\mathbf{p}_n$. If, on the other hand, vortices are tangled, then the vector $\boldsymbol{\omega}$ alone does not carry sufficient information to describe the vortex arrangement. How to obtain the mutual friction in such a case is still an open problem, even though some progress can be made by using concepts of quantum turbulence imported from the study of vortex tangles in superfluid $^4$He~\cite{andersson_turbulence,mongiovi2017MNRAS,Haskell_barenghi_2020}. 

Now, we can derive the phenomenological form of $\mathbf{F}^{n}$, which must transform like $\dot{\mathbf{p}}_n$ under a Galilean transformation. Since we have at our disposal $\hat{\boldsymbol{\omega}}$ and the relative velocity  $\mathbf{v}_{np}=\mathbf{v}_n-\mathbf{v}_p$, we can always project $\mathbf{F}^{n}$ on a right-handed orthogonal basis: 
\begin{equation}
   \perp_{\hat{\boldsymbol{\omega}}} \mathbf{v}_{np}=  -\hat{\boldsymbol{\omega}}\times(\hat{\boldsymbol{\omega}}\times\mathbf{v}_{np}) \, ,
    \qquad 
    \hat{\boldsymbol{\omega}}\times\mathbf{v}_{np} \, ,
    \qquad 
    (\hat{\boldsymbol{\omega}} \cdot \mathbf{v}_{np}) \hat{\boldsymbol{\omega}} 
    \label{pizzofranco}
\end{equation}
where the operator $\perp_{\hat{\boldsymbol{\omega}}}$ is the projector that kills the components parallel to $\hat{\boldsymbol{\omega}}$. Therefore, the force per unit volume $\mathbf{F}^{n}$ can be decomposed as
\begin{equation}
   \mathbf{F}^{n}/(\rho_n \, | \boldsymbol{\omega} |)= 
     \mathcal{B}_d \, \hat{\boldsymbol{\omega}}\times(\hat{\boldsymbol{\omega}}\times\mathbf{v}_{np})
   +\mathcal{B}_c \, \hat{\boldsymbol{\omega}}\times\mathbf{v}_{np}
   +\mathcal{B}_k \, \hat{\boldsymbol{\omega}} (\hat{\boldsymbol{\omega}} \cdot \mathbf{v}_{np}) 
   \, ,   
   \label{MF_pheno}
\end{equation}
where the factors $\rho_n$ and $ | \boldsymbol{\omega} | $ have been introduced to make the $\mathcal{B}$ coefficients dimensionless. The three subscripts $d$, $c$ and $k$ stand for ``dissipative'', ``conservative'' and ``Kelvin''. 
If we want  $\mathbf{F}^{n}$ to be a force that drives the system towards the zero-lag state $\mathbf{v}_{np}=0$, then we have to require that the dissipative part is directed in the opposite direction to $\perp_{\hat{\boldsymbol{\omega}}} \mathbf{v}_{np}$, so that $\mathcal{B}_d>0$. Similarly, we can conclude that $\mathcal{B}_k<0$. 
In this way, the entropy of the fluid is guaranteed to satisfy the second law, so that the homogeneous zero-current state is a stable equilibrium~\cite{GusakovHVBK,GAVASSINO_iordanskii}.  

The $\mathcal{B}_k$ coefficient may have a role when the vortices develop Kelvin waves (helical displacements of a vortex's core), as discussed by \citet{barenghi_MF_1983}. It is customary to set $\mathcal{B}_k=0$ in glitch and $^4$He studies but it should be included in the case a certain amount of quantum turbulence is expected\footnote{
    The phenomenological expression \eqref{MF_pheno} is general enough that it may be valid also if vortices are not parallel to each other in the fluid element. 
    In particular, \eqref{MF_pheno} also contains the  \citet{GorterMellink} result for the mutual friction in the isotropic turbulent regime as the limiting case $\mathcal{B}_c=0$, $\mathcal{B}_d=-\mathcal{B}_k \propto |\mathbf{v}_{np}|^2$. 
    The difference with the non-turbulent case is that $\boldsymbol{\omega}$ alone is not sufficient to describe the vorticity in the fluid element, so more involved additional terms may be needed in \eqref{MF_pheno}.}. 
 Relativistic extension of the geometric construction in \eqref{pizzofranco} can be found in the seminal work of \citet{langlois_etal98} and~\cite{ander_MFGR,GusakovHVBK,gavassino2020MNRAS,GAVASSINO_iordanskii}. 

Finally, the phenomenological $\mathcal{B}$ coefficients in \eqref{MF_pheno} can be functions of the lag  $\mathbf{v}_{np}$ and the vorticity, making the mutual friction nonlinear in the lag \cite{GorterMellink,andersson_turbulence,antonelli17,celora2020MNRAS}. 
The hard task is to find their expressions from the physics at the microscopic and mesoscopic scale.

\subsection{The relation between mutual friction and vortex dynamics}

The three $\mathcal{B}$ coefficients in \eqref{MF_pheno} have been introduced on the basis of a purely geometrical argument. We can address the problem of their determination by assuming a certain phenomenological model for the dynamics of vortex lines. This will not solve all the problems, as the vortex's dynamics will be governed by a new set of phenomenological parameters (again, introduced geometrically), that should then be derived from the microscopic properties of the system at the scale of the vortex's core. However, this allows us to go one step down the staircase of physical scales (from the macroscopic scale to the mesoscopic scale).
In this way, we can also clarify the relation between the average vortex velocity and the mutual friction \cite{Sedrakian_precession_1999,andersson_MF,antonelli2020MNRAS}. 
For locally parallel vortex lines, we can introduce the average vortex velocity $\mathbf{v}_L$ by writing down the transport equation for the macroscopic vorticity $\boldsymbol{\omega}$ field,
\begin{equation}
\partial_t \boldsymbol{\omega}  + \nabla \times( \boldsymbol{\omega}  \times \mathbf{v}_L )  \, = 0  \, .
\label{hvbkn_curl2}
\end{equation}
In the above equation,  $\mathbf{v}_L$ can be taken to be orthogonal to $\boldsymbol{\omega} $ with no loss of generality. In this way, the current of vortex lines in the fluid element is simply $n_v \mathbf{v}_L$. A rigorous geometrical definition of $\mathbf{v}_L$ in General Relativity is given in \cite{gavassino2020MNRAS}.
Equation \eqref{hvbkn_curl2} must also be consistent with the vorticity equation derived by taking the curl of  \eqref{eulerox} for $x=n$, 
\begin{equation}
\partial_t \boldsymbol{\omega}  + \nabla \times( \boldsymbol{\omega}  \times \mathbf{v}_n  )  
= \nabla \times (\rho_n^{-1}\, \mathbf{F}_{n}) \, . 
\label{hvbkn_curl}
\end{equation}
Equations \eqref{hvbkn_curl2} and \eqref{hvbkn_curl} are consistent if
\begin{equation}
\mathbf{F}_{n}  =  -\rho_n  \boldsymbol{\omega} \times( \mathbf{v}_L - \mathbf{v}_n ) = 
-\rho_n |\boldsymbol{\omega}| \mathbf{f}_M \, .
\label{Fn_general}
\end{equation}
This tells us that the force between the components is zero when the vortices are advected in such a way that $\mathbf{v}_L = \mathbf{v}_n$. In the above equation we  also defined the quantity $\mathbf{f}_M$, the \emph{Magnus force} acting on a straight vortex segment\footnote{
    The quantity $\mathbf{f}_M =\boldsymbol{\omega} \times( \mathbf{v}_L - \mathbf{v}_n )$ is just a rescaling of what in the literature is sometimes called ``Magnus force per unit length'' (of vortex line), i.e. $\rho_n\boldsymbol{\omega} \times( \mathbf{v}_L - \mathbf{v}_n )$. 
    Apart from the overall sign, $\mathbf{f}_M $ is the same quantity appearing in \eqref{MF_pheno}. 
    }. 
From \eqref{Fn_general}, we see that if vortices stick to the normal component because of pinning, then $\mathbf{v}_L = \mathbf{v}_p$ and $\mathbf{F}_{n} = \rho_n  \boldsymbol{\omega} \times  \mathbf{v}_{np}$, namely $\mathcal{B}_c = 1$ and $\mathcal{B}_d=0$: there is no dissipation in this limit. More generally, if we have a model to calculate the average vortex velocity in a fluid element $\mathbf{v}_L$ for any given value of the lag $\mathbf{v}_{np}$, then we can substitute the function $\mathbf{v}_L(\mathbf{v}_{np})$ back into \eqref{Fn_general} and obtain the expression of the $\mathcal{B}$ coefficients in \eqref{MF_pheno}, as we show below with a simple model.
\\
\\
\indent
\emph{\textbf{Magnus and drag forces} -} In both $^4$He \cite{barenghi_MF_1983} and NSs \cite{mendell1991ApJ_II,andersson_MF}, the  simplest phenomenological model for vortex dynamics is in terms of balance of the Magnus force and a drag force -- proportional to a dimensionless drag parameter $\mathcal{R}$ -- acting on a straight segment of vortex line,
\begin{equation}
    \mathbf{f}_M 
    - \mathcal{R} \perp_{\hat{\boldsymbol{\omega}}} (\mathbf{v}_L - \mathbf{v}_p) 
    = 
    \hat{\boldsymbol{\omega}} \times \left[ \mathbf{v}_L - \mathbf{v}_n  
    + \mathcal{R}  
      \hat{\boldsymbol{\omega}} \times ( \mathbf{v}_L - \mathbf{v}_p) \right] = 0 \, .
      \label{rollino}
\end{equation}
The above equation can be inverted in the plane orthogonal to $\hat{\boldsymbol{\omega}}$ to find 
$ \mathbf{v}_L$ as a function of $\mathbf{v}_n$ and $\mathbf{v}_p$. Plugging the result for $ \mathbf{v}_L$ back into \eqref{Fn_general} allows us to find (we can assume $\mathcal{B}_k =0$):
\begin{equation}
\label{eq:BcBd}
    \mathcal{B}_c \, = \, \frac{\mathcal{R}^2 }{ 1+ \mathcal{R}^2} 
    \qquad \qquad
    \mathcal{B}_d \, = \, \frac{\mathcal{R} }{ 1+ \mathcal{R}^2}\,.
\end{equation}
This tells us that, as long as $\mathcal{R}$ can be treated as a constant, the mutual friction is linear in the lag $\mathbf{v}_{np}$. 

Now, the physical problem is to understand which microscopic processes have to be taken into account to estimate the mesoscopic parameter $\mathcal{R}$. 
In the outer core, it is believed that the presence of entrainment causes the magnetisation of the neutron superfluid vortices \cite{alpar84rapid,andersson_MF}: the current of protons that are entrained by the local circulation of a vortex creates a sort of solenoid. The scattering of the relativistic free electrons with these magnetised vortex cores is the main process that is taken into account in the estimates of $\mathcal{R}$.
In the inner crust, the parameter  $\mathcal{R}$ probably depends also on the relative velocity between the superfluid vortices and the normal component, meaning that the drag is nonlinear~\cite{celora2020MNRAS}. For low values of this relative velocity the drag is caused by phonon excitation in the crustal lattice \cite{Jones1990}, while for higher values of this parameter, the drag force is caused by excitation of Kelvin waves on the superfluid vortices~\cite{Jones1992, epstein_baym92, graber+2018}.

It is also worth mentioning that, in the inner crust, different chemical gauge choices -- different definitions of the neutrons that are considered free \cite{carter_macro_2006} -- demand a more general treatment than the one outlined above: in addition to the Magnus and the drag forces,  also other contributions could be included in the vortex equation of motion \eqref{rollino}.
This is done to ensure that the final hydrodynamic theory is well behaved under redefinitions of the density $\rho_n$ of the n-component~\cite{GAVASSINO_iordanskii}. 
\\
\\
\indent
\emph{\textbf{Magnus, drag and pinning forces} -} If vortices are immersed in a non-homogeneous medium, then their interaction with the substrate must be taken into account. The vortex creep model of \citet{Alpar1984a} deals with this problem by assuming that the dependence of the jump rate from one pinning center to another one is given by the typical  Arrhenius formula, where the activation energy for the jump is corrected by the presence of the background superfluid flow. This makes  $\mathbf{F}_{n}$ nonlinear in the velocity lag \cite{guerci_2017MNRAS,celora2020MNRAS}. 
Moreover, the simple model defined by \eqref{rollino} is valid for \emph{locally} straight and parallel vortices: the fact that vortices can bend in the fluid element -- because of excitations or due to the interactions with the crustal impurities \cite{link2009,Wlazlowski2016} -- should also be taken into account in more realistic extensions of the model.

When an additional pinning force is included in equation \eqref{rollino} for the motion for the vortices (as done in, e.g., \cite{sedrakianRepinning,link2009,haskell15,Wlazlowski2016}), the $\mathcal{B}$ coefficients are not constant anymore, giving rise to a nonlinear force even if $\mathcal{R}$ is an assigned constant parameter \cite{antonelli2020MNRAS}. 
For small  $\mathbf{v}_{np}$ the parameter $\mathcal{B}_d$ is highly suppressed, while $\mathcal{B}_c\approx 1$. When the lag reaches a certain value (dependent on the details of the pinning landscape), the vortices start to move and $\mathcal{B}_d$  tends to grow. For very high values of the lag, the pinning interaction will be just a small correction to the Magnus and drag forces, so we can expect that the linear regime in \eqref{eq:BcBd} is eventually recovered for very high values of the lag (remember, however, that the parameter $\mathcal{R}$ may not be constant in the first place). 

Pinning can also induce \emph{hysteresis} in the pinning-unpinning transition as the lag is first increased and then decreased in a cyclic transformation, making the behaviour of the parameters $ \mathcal{B}_c $ and $ \mathcal{B}_d $ history-dependent \cite{antonelli2020MNRAS}. 
As discussed in \cite{antonelli2020MNRAS}, this hysteresis mechanism can provide a natural explanation for the behaviour of most pulsars, which is probably neither time-predictable nor slip-predictable (see Fig.~\ref{fig:quake}).

\subsection{Circular models}
\label{circo}

In general, the fluid equations \eqref{eq:continuity} and \eqref{eulerox} can be solved only numerically \cite{Peralta2005ApJ,Howitt2016}. Thus, to obtain a glitch model that is more easily tractable, it is necessary to simplify the problem by considering only a specific subset of all possible fluid motions. 
We assume a circular (not necessarily rigid) motion and derive the equations for the angular velocity of the two components. This is mostly a pedagogical exercise, useful to gain some understanding of the backbone of many glitch models. 
Moreover, this exercise also sheds some light on the body-averaging procedure often invoked when dealing with rigid components, see Sec.~\ref{sec1.1}. 

We use standard cylindrical coordinates $(x, \varphi, z)$, where $x$ is the distance from the rotation axis of the star; the equatorial plane is  $z=0$. 
In the absence of precession and neglecting meridional circulation\footnote{
    In General Relativity this would give rise to a ``circular spacetime'' \cite{Bonazzola_1993,Andersson2000,gavassino2020MNRAS}.
    Quasi-circular models with meridional circulation -- a kind of motion like the one in convection cells -- are discussed in~\cite{sourie2021MNRAS}.
}, we have
\begin{equation}
\label{ansatz}
\mathbf{v}_n   = x \, \left(\Omega_p(x,z,t)+\Omega_{np}(x,z,t) \right)\, \mathbf{e}_\varphi 
\qquad \qquad  \quad
\mathbf{v}_p   = x \, \Omega_p(x,z,t)\, \mathbf{e}_\varphi\, .    
\end{equation}
It is also useful to define an auxiliary angular velocity variable directly from the momentum $\mathbf{p}_n$,   
\begin{equation}
    \Omega_v(x,z,t) =p_n^\varphi/x =  \Omega_p+ (1-\epsilon_n)\Omega_{np}\, .
    \label{eq:omegav}
\end{equation}
Therefore, the coarse-grained vorticity \eqref{asdrubale} reads
\begin{equation}
 \boldsymbol{\omega} 
 \, = \, 
 -x \partial_z \Omega_v \, \mathbf{e}_x 
 \,+ \,
   (2\Omega_v + x \dx \Omega_v)\, \mathbf{e}_z \, ,
   \label{localFO}
\end{equation}
that reduces to the standard results $ \boldsymbol{\omega} =2\Omega_v \, \mathbf{e}_z$ and $n_v = 2 \Omega_v/\kappa$ if the superfluid is rigidly rotating at the macroscopic scale. 
We can use \eqref{localFO} and \eqref{MF_pheno} to obtain\footnote{
    Equation \eqref{werrrro} assumes locally parallel vortices ($\mathcal{B}_k=0$). However, vortices do not necessarily have to be straight, as we have differential non-uniform rotation, but microscopic vorticity follows macroscopic vorticity lines \cite{gavassino2020MNRAS}. 
    If no vortex reconnection is allowed, this can lead to large-scale turbulence as vortices may wrap around the rotation axis~\cite{greenstein70}. This would also break the working assumption of negligible meridional circulation.
} 
\begin{equation}
\label{werrrro}
    \mathbf{F}_n  = \rho_n\, x\, \Omega_{np} \Big( \mathcal{B}_c  \left[ x \partial_z \Omega_v \mathbf{e}_z + \left(2 \Omega_v + x \partial_x \Omega_v \right) \mathbf{e}_x \right]  
                - \mathcal{B}_d\, |\boldsymbol{\omega}|\, \mathbf{e}_\varphi \Big)\, .
\end{equation}
We are interested in specifying the motion along the azimuthal $\varphi$ direction. 
In fact, the components of  \eqref{eulerox} along $\mathbf{e}_x$ and $\mathbf{e}_z$ contain no time derivative: given the restrictive ansatz in \eqref{ansatz}, their role is to define the instantaneous configuration of the rotating NS, whose deviation from the spherical hydrostatic equilibrium must be constantly adjusted as $\Omega_n$ and $\Omega_p$ vary in time. 
In the slowly rotating limit \cite{andersson_comer01}, it is not a bad approximation to assume some given spherical stratification of the star, 
\begin{equation}
\rho_{n,p}(t,x,z)\approx\rho_{n,p}(r)\, , 
\qquad \text{where} \qquad
r=\sqrt{x^2+z^2}\, .
\end{equation}
The same is assumed to be valid also for the entrainment parameters $\epsilon_{n,p}$.
The dynamics along  $\mathbf{e}_\varphi$ is particularly simple since the Lie derivative and the gradient term do not have any component directed along $\mathbf{e}_\varphi$:
\begin{align}
\label{eq:first}
	& x \rho_n  (\partial_t \Omega_n - \epsilon_n \partial_t \Omega_{np}) = 
	 x \rho_n \partial_t \,\Omega_v \, = \, {F}^\varphi_n
	\\
\label{eq:second}
	& x \rho_p (\partial_t \Omega_p + \epsilon_p \partial_t \Omega_{np})\, = \, -{F}^\varphi_n  + F^\varphi_{Bp}
	\\
	& {F}^\varphi_n = -\rho_n x \mathcal{B}_d |\boldsymbol{\omega}| \Omega_{np}
    \\
	& |\boldsymbol{\omega}| = \sqrt{(2 \Omega_v + x \partial_x \Omega_v )^2 + (x \partial_z \Omega_v)^2 }\, ,
	\label{omegaxz}
\end{align}
where we have also added an extra external force $F^\varphi_{Bp}$ to model the braking effect of dipole radiation on the normal component. An external torque due to gravitational wave emission would act on both components, see e.g.~\cite{Meyers2021MNRAS}. 
Note that, despite $\mathcal{B}_c\neq 0$, this parameter only appears in the two neglected equations that define the stellar structure: perfect pinning ($\mathcal{B}_d=0$, $\mathcal{B}_c=1$) can strain a rigid crust~\cite{Ruderman1991_Tectonics_I,giliberti2020MNRAS}.

If we sum \eqref{eq:first} and \eqref{eq:second} and define the total density $\rho = \rho_n+\rho_p$, we have:
\begin{equation}
    x ( \rho \,\partial_t \Omega_p \, + \, x \rho_n \partial_t \Omega_{np})  
    \, = \,  
    {F}^\varphi_{Bp} \, .    
    \label{piazzolla}
\end{equation}
that, in the limit ${F}^\varphi_{Bp}=0$ is the $\varphi$ component of $\eqref{prezzemolo}$. 
The above equation may be used in place of, say,~\eqref{eq:second}.

Another widely used assumption in glitch modelling is that the normal component rotates rigidly. This assumption is beyond the fluid equations \eqref{eulerox}, as it is motivated by physics  (normal viscosity, magnetic field and elasticity) that is not implemented in the very incomplete model we are using here. If we simply put $\Omega_p=\Omega_p(t)$ into \eqref{eq:first} and \eqref{eq:second}, then we end up with an inconsistent set of equations.  
Assuming rigid motion for the normal component forces us to substitute the local equation \eqref{piazzolla} with an integral one,
\begin{equation}
\label{azzo}
\int d^3x \,x^2 \,(\rho\, \partial_t \Omega_p \, + \, \rho_n\, \partial_t \Omega_{np}) 
\, =\, 
\int d^3x \,x \,\,  {F}^\varphi_{Bp} 
\qquad [\,d^3x=2\pi x dxdz\,] .
\end{equation}
It is only at this point that  we can really restrict the motion of the p-component to be rigid: assuming $\Omega_p=\Omega_p(t)$ allows us to bring it outside the integral to obtain  
\begin{equation}
I \, \dot{\Omega}_p \, + \, \int d^3x \, x^2 \rho_n \partial_t\Omega_{np} \, =\, 
-I \oi \, ,
\label{eq:rigid}    
\end{equation}
where  $I$ is the total moment of inertia of the star and $\oi>0$ is a phenomenological parameter -- or a function of time and/or the angular velocities -- that sets the strength of the total braking torque. Practically, $\oi$ is just another way of writing the (already phenomenological and unspecified) ${F}^\varphi_{Bp}$ term in \eqref{azzo}. 
% \begin{equation}
% \oi \, = \, \frac{1}{I} \, \left| \,  \int d^3x \,x \,\,  {F}^\varphi_{Bp} \, \right| \,.
% \end{equation}
The other equation that we need can be taken to be \eqref{eq:first}:
\begin{equation}
\partial_t\, \Omega_v(t,x,z) \, = \, - \mathcal{B}_d \, | \boldsymbol{\omega} | \, \Omega_{np} \, 
\, = \, - \mathcal{B}_d \,| \boldsymbol{\omega} | \, \frac{\Omega_{vp}}{1-\epsilon_n} \, ,
\label{bhubhu}
\end{equation}
where we recall that, $\Omega_{vp}= \Omega_v- \Omega_p = (1-\epsilon_n) \, \Omega_{np}$
is the local lag between the superfluid\footnote{
    The variable $\Omega_v$ is defined by the vortex configuration and is, basically, the \emph{superfluid velocity} of Tisza and Landau (not to be confused with the \emph{velocity of the species that is superfluid}, $\Omega_n$). Hence, $\Omega_v$ is not an exotic variable but is the standard choice for laboratory superfluids. It has also the nice property of being chemical gauge independent \cite{carter_macro_2006,GAVASSINO_iordanskii}.
} angular velocity $\Omega_v$ and the one of the normal component.
If we want to use $\Omega_v$ in place of $\Omega_n$ as primary variable, then \eqref{eq:rigid} can be equivalently written as
\begin{equation}
I \dot{\Omega}_p \, + \, \int d^3x \, x^2 \, \rho^*_n \, \partial_t\Omega_{vp} \, =\, 
-I \oi \, ,
\label{eq:rigid2}    
\end{equation}
where $\rho^*_n$ is the superfluid density introduced in \eqref{marzullo}. 
In this way, the model is formulated in terms of the chemically gauge invariant quantities
$\Omega_p$, $\Omega_v$, $\rho^*_n$ and $\rho$. 
\\
\\
\indent
\emph{\textbf{Axially symmetric dynamical equations} -} To summarize, for a given static and spherical stratification defined by $\rho_{n,p}(r)$ and $\epsilon_{n}(r)$, the axially symmetric equations are \eqref{eq:rigid} and \eqref{bhubhu}.
The two equations are more conveniently written by selecting $\Omega_p(t)$ and $\Omega_v(t,x,z)$ as primary variables, together with $\rho^*_n$ and $\rho$ for the densities. 
Introducing an average operation  as
\begin{equation}
     dI_v = d^3x \, x^2 \, \rho^*_n  
     \qquad \quad I_v = \int dI_v 
     \qquad \quad \langle \, f \, \rangle = I_v^{-1} \! \! \int dI_v \, f(t,x,z)  \,  ,
     \label{crispino}
\end{equation}
the system defined by \eqref{eq:rigid2} and \eqref{bhubhu} is equivalent to   
\begin{align}
 & I \partial_t\,{\Omega}_p(t) \, + \,I_v \, \partial_t \langle \, \Omega_{vp} \, \rangle  \, =\, -I \oi  
 \label{qser}
\\
 & \dt \Omega_v(t,x,z) \, =  \, - \mathcal{B}_d(r,\Omega_{vp}) \,| \boldsymbol{\omega} | \, \frac{\Omega_{vp}(t,x,z)}{1-\epsilon_n(r)}
 \label{omVeq}
\end{align}
where  $|\boldsymbol{\omega}| $ is given in \eqref{omegaxz}, the local lag is $\Omega_{vp}(t,x,z)=\Omega_v(t,x,z)-\Omega_p$ and the averaged lag $\langle\Omega_{vp}\rangle=\langle\Omega_v\rangle-\Omega_p$ is a function of time only.
Since the stratification is given, the drag parameter $\mathcal{B}_d$ can be assumed to be fixed and almost spherically stratified, i.e. $\mathcal{B}_d \approx \mathcal{B}_d (r) $, or to contain also an extra dependence on the lag $\Omega_{vp}$, see the discussion below \eqref{MF_pheno}. Since it $\mathcal{B}_d $ is, ultimately, a phenomenological function of the lag, the explicit entrainment correction in \eqref{omVeq} may be adsorbed directly in it: the meaning of the explicit entrainment in \eqref{omVeq} is just to remind us that $\mathcal{B}_d $ is not chemical gauge invariant~\cite{GAVASSINO_iordanskii}.

Moreover, in order to derive \eqref{qser} and \eqref{omVeq}, we have used only the momentum equations \eqref{eulerox} because the conservation equations \eqref{eq:continuity} are automatically satisfied when assuming axial symmetry and no meridional flows. 

Finally, \eqref{qser} and \eqref{omVeq} have been derived in the limit of ``slack'' vortices, in which the small energy cost (tension) of bending a quantum vortex is neglected \cite{antonelli_GR_2018}. 
This effect of ``vortex tension'' is sometimes included in extensions of the HVBK equations \cite{Qbook}. When this vortex tension \cite{andersson_turbulence,link14,haskell15} is taken into account, then an additional term containing derivatives along $z$ and $x$ should appear in the left-hand side of \eqref{omVeq}. 

Opposite to the slack limit, we have the extreme scenario in which $\Omega_{v}$ is cylindrical (no $z$ dependence, so that $\boldsymbol{\omega}$ is everywhere parallel to the rotation axis), which is formally recovered in the limit where the vortex tension is infinite~\cite{antonelli17}. 
This scenario, despite being unrealistic
%\footnote{
%    For example, one should be careful not to rule out a priori turbulent motion and its consequences on mutual friction: this is a limit of oversimplified cylindrical or rigid models to be kept in mind -- see the discussion in~\cite{antonelli17,Levin2023arXiv230611775L}.
%    }
, is implicitly used in all studies based on cylindrical models -- e.g., \cite{Alpar1984a,link1996,pizzochero2011,haskell+2012,graber+2018,khomenko2018,haskell2018MNRAS,erbil_vela_2020MNRAS} -- because it further simplifies the hydrodynamic equations. 
\\
\\
\indent
\emph{\textbf{Models with rigid components} -} The body-averaged model defined by \eqref{baym1} and \eqref{baym2} lacks direct connection with the local properties of neutron star matter. 
However, it is possible to link its parameters with the quantities appearing in \eqref{qser} and \eqref{omVeq}. 
The rigid model can be recovered by averaging \eqref{omVeq} over by using \eqref{crispino},
\begin{equation}
 \langle \partial_t \Omega_v \rangle \, =  \, \partial_t \langle  \Omega_v \rangle 
 \, =  \, - \left\langle \mathcal{B}_d  \,| \boldsymbol{\omega} | \, \frac{\Omega_{vp}}{1-\epsilon_n} \right\rangle
 \label{pallone}
\end{equation}
Clearly, the approximate reduction to a rigid model works well if spatial correlations between the different quantities are weak, that is 
\begin{equation}
    \left\langle \mathcal{B}_d \,| \boldsymbol{\omega} | \, \frac{\Omega_{vp}}{1-\epsilon_n} \right\rangle
    \approx 
    \left\langle \, \frac{\mathcal{B}_d  \,| \boldsymbol{\omega} | }{1-\epsilon_n} \right\rangle
    \langle  \Omega_{vp} \rangle 
    \approx 
    2 \langle  \Omega_{v} \rangle  \left\langle \, \frac{\mathcal{B}_d  \,  }{1-\epsilon_n} \right\rangle
    \langle  \Omega_{vp} \rangle \, .
    \label{eq:correlations_weak}
\end{equation}
The factor $\langle  \Omega_{v} \rangle\approx \Omega_p$ is almost constant, but the above expression can still be highly nonlinear if $\mathcal{B}_d$ depends on the local lag. The linear friction model of \citet{BAYM1969} can only be found if $\mathcal{B}_d$ is locally constant over the timescale of interest. In this case, a direct comparison of \eqref{qser} and \eqref{pallone} with \eqref{baym1} and \eqref{baym2} allows us to make the following identifications:
\begin{equation}
\begin{split}
        & I_{s} = I_v \qquad  \qquad \, \, I_{o} = I-I_v  
    \\  & \Omega_s = \langle \Omega_v \rangle  \qquad \quad   \Omega_o = \Omega_p 
    \\  & \frac{1}{\tau} = \frac{I-I_v}{I \, \tau_{\rm{r}} }\, \approx \, 2  \Omega_{p}   \left\langle \, \frac{\mathcal{B}_d  \,  }{1-\epsilon_n} \right\rangle
\end{split}
    \label{tau_rigid}
\end{equation}
For example, this tells us how a fit of the timescale $\tau_{\rm{r}}$ from observations can be interpreted in terms of the physical input at smaller scales. General relativistic corrections should be also taken into account, which affect both the expressions for the moments of inertia and for $\tau$ \cite{antonelli_GR_2018,sourie_etal16,gavassino2020MNRAS}. 

A similar reduction to a rigid model can be performed by using $\Omega_n$ instead of $\Omega_v$. In this case, extra internal torques due to the entrainment coupling appear in the final equations~\cite{chamel_carter_2006MNRAS,SP10}: this is also the reason why entrainment corrections enter in slightly different ways in theoretical upper limits to a pulsar's activity, cf. \eqref{eq:constraint_activity} and~\eqref{eq:constraint_activity2}. 
Finally, the superfluid region (the domain of $\Omega_v$) can also be split into an arbitrary number of different layers to obtain the multi-component generalization of this minimal two-component model~\cite{montolimagistrelli+2020}.

\section{Dynamical phases and the Vela's 2016 glitch precursor}
\label{sec:dynamical_phases}
It is interesting to make some general considerations on the dynamics defined by \eqref{qser} and \eqref{omVeq}. Clearly, \eqref{qser} and \eqref{omVeq} do not allow tracking the stresses in the solid components: possible starquakes or slow changes of the moments of inertia $I$ and $I_v$ should be implemented by hand, as in Sec.~\ref{sec:2c}. Hence, we assume here the \citet{ANDERSON1975} paradigm, in which the moments of inertia $I$ and $I_v$ can be taken constant. 

The angular momentum balance \eqref{qser} and the friction equation \eqref{omVeq} have two different roles. In particular, the momentum balance \eqref{qser} allows us to identify some dynamical phases sketched in Fig.~\ref{fig:phases}, while the possibility to switch from one phase to another depends on the details of friction, i.e., on the details of the $\mathcal{B}_d(\Omega_{vp})$ function or, more generally, on the statistical properties of the landscape in Fig.~\ref{fig:pinningladscape}. 
In brief, the possibility of exploring certain phases depends on how the angular momentum can be exchanged between the components and transported across different regions during the pulsar's life. 

Since there is still no complete understanding of these issues related to friction and history dependence, it may be instructive to consider each dynamical phase as a separate (theoretical) possibility. 
The following analysis is valid for a system where the superfluid component can rotate non-uniformly: this system has an infinite number of degrees of freedom, so that each phase in Fig. \ref{fig:phases} is highly degenerate (it can be realized in a number of ways). The same reasoning can be applied to simpler models with two or more rigid components, that have a finite number of degrees of freedom and are defined by ordinary differential equations.
  
We plot the different phases defined by the angular momentum balance \eqref{qser} in Fig.~\ref{fig:phases}.
\begin{figure}
  \begin{minipage}[c]{0.7\textwidth}
    %\centering
    \includegraphics[width = 0.95\textwidth]{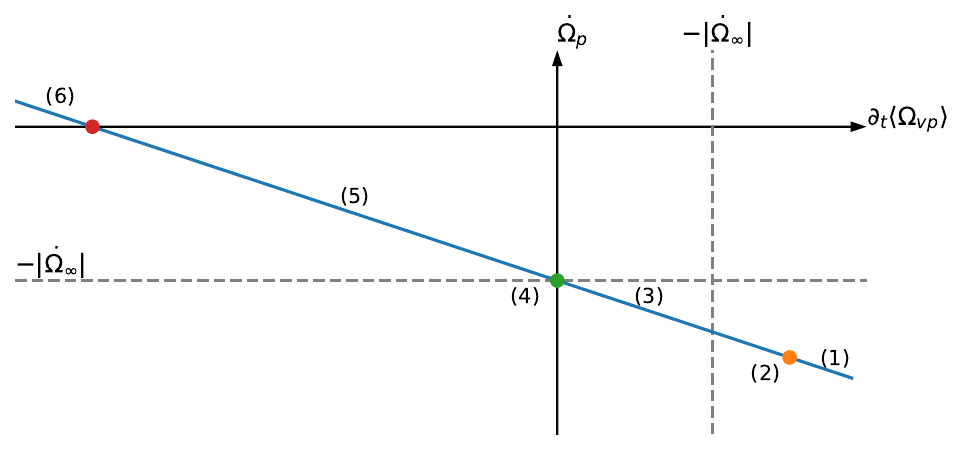}
          \end{minipage} %\hspace{2mm}
      \begin{minipage}[c]{0.29\textwidth}
    \caption{
    The evolution of the total angular momentum \eqref{qser} can be regarded as a constraint (the blue line) in the 
    $( \partial_t \langle {\Omega}_{vp} \rangle ,\partial_t {\Omega}_p)$ plane, whatever the number of internal ``components'' (formally infinite if the superfluid can rotate non-uniformly) or the details of the mutual friction. 
    The numbers indicate the different dynamical phases described in the main text. The slope of the curve depends on the constant ratio $I_v/I$. }
    \label{fig:phases}
    \vspace{4mm}
     \end{minipage}
\end{figure}
Going from negative to positive values of $\partial_t\Omega_p$ we have: 

\begin{enumerate}

\item \emph{Inward vortex creep -} From \eqref{qser} we see that $\dt \Omega_p < - I/(I-I_v)\oi$ can be realized only if the internal torque contributes to slowing down the observable component (the internal torque has the same sign as the external one). Since  $\mathcal{B}_d>0$, as demanded by the second law of thermodynamics \cite{GusakovHVBK,GAVASSINO_iordanskii}, this phase can be realized only if there is a sufficiently extended region in which the lag is negative (recall that $\text{sign} \ovp = \text{sign} \Omega_{np}$). This also implies that the vortex velocity is, on average, directed inward.\\

\item \emph{Average pinning -} In this phase, the average superfluid momentum is conserved, $\partial_t \langle \Omega_v\rangle =0$. 
The lag between the two components builds up as $\partial_t \langle \ovp \rangle = -\partial_t \Omega_p$, with $\dt \Omega_p = -I/(I-I_v) \oi$ from equation \eqref{eq:rigid2}. This phase can be achieved with perfect pinning (the superfluid is \emph{completely decoupled} from the normal component) or by a balance of inward and outward creep in different regions. \\

\item \emph{Moderate outward creep -} The average lag satisfies $0 < \partial_t \langle \ovp \rangle < -\partial_t \Omega_p$, meaning that the superfluid is only partially decoupled.
This state is likely to occur during the slow post-glitch relaxation, when the post-glitch lag is expected to slowly increase because of gradual repinning, $\dt \langle\ovp\rangle >0$.\\

\item \emph{Steady-state -} The body-averaged lag is constant in time, $\partial_t \langle \ovp \rangle = 0$ and $\dt \Omega_p= \dt \langle \Omega_v  \rangle= -\oi$. However, the local lag $ \ovp$ can fluctuate, for example, if a sequence of packets of unpinned vortices creeps outward.\\

\item \emph{Speed up -} This phase is defined as $-\oi<\dt \Omega_p<0$: the observable component is spinning down but the average lag is decreasing in such a way that $-I/I_v\oi <\dt \langle \ovp \rangle < 0$. The normal component receives some extra angular momentum from the superfluid but at such a low rate that it is still spinning down. This phase could have been observed in the Crab pulsar as the result of a small increase in the Crab pulsar's internal temperature \cite{Vivekanand2017A&A}, which has the effect of increasing the outward vortex creep rate~\cite{Alpar1984a}.\\

\item \emph{Spin-up -} The observable component spins up, $\partial_t \Omega_p>0$ and the lag is decreasing fast $\dt \langle \ovp \rangle < - I/I_v \oi$.
The inequality is stronger in real glitches, $\dt \langle \ovp \rangle \ll -I/I_v \oi$. However, a slow spin-up event has also been observed in the Crab pulsar following a much faster, unresolved, jump \cite{shaw2018_largestCrab}. 
The particular value $\partial_t \Omega_p = 0$ can be realized if the internal torque exactly balances the external one. This should happen for a brief moment when the observable component reaches the maximum amplitude in a fast glitch (e.g., during an ``overshoot'', see Sec.~\ref{sec:evolution}), or in a slower spin-up event like the one in the Crab.
\end{enumerate}
This analysis is independent of the details of the mutual friction, apart from the thermodynamic requirement $\mathcal{B}_d>0$, but it depends on the fact that the moments of inertia are assumed to be constant. 
\\
\\
\indent
\emph{\textbf{The precursor in Vela's 2016 glitch} -} \citet{ashton+2019} inferred a peculiar feature in the Vela pulsar's phase residuals around the glitch recorded by \citet{palfreyman+2018}. 
This feature may be interpreted as a spin-down (an anti-glitch), occurring just before the main spin-up event, and is possibly linked to the triggering of the glitch.
Assuming that this behaviour could be explained in terms of angular momentum exchange between the superfluid and the normal component, it is interesting to try to understand to which dynamical phase this spin-down ``precursor'' may belong to. 

According to the fit performed in \cite{ashton+2019}, this sudden decrease of the Vela's rotational frequency ($\nu \approx 11\,$Hz) has an amplitude $\Delta \nu \sim -5 \times 10^{-6}\,$Hz and it is likely to occur over a time span of $\Delta t\sim 100$s, implying $\dot{\Omega}_p \approx 2 \pi \Delta \nu/\Delta t \sim -10^{-7}\,$rad/s$^2$. 
Since a spin-down of the normal component would increase the velocity lag, this precursor may also trigger the subsequent glitch by causing a critical lag with the superfluid \cite{ashton+2019}.

We should compare the above estimate of $\dot{\Omega}_p$ with the secular spin-down parameter of the Vela, $|\dot{\Omega}|_\infty  \approx 10^{-10}$rad/s$^2$: by looking at Fig. \ref{fig:phases}, we see that the Vela should be in either phase 1, 2 or 3 during the anti-glitch precursor. 
Let us assume that phase 2 is realized, which implies complete decoupling of the superfluid: the observable component could spin down at a faster rate as the external torque acts on a reduced moment of inertia, especially if $I_v \approx I$, which is the extreme situation where the NS is almost entirely superfluid and all this superfluid is practically decoupled (or there is a balance between inward and outward creep in different regions). 
Assuming phase 2, we have  $-\dot{\Omega}_p/|\dot{\Omega}|_\infty =  I/(I-I_v) \approx  10^3$, so that the blue line in Fig.~\ref{fig:phases} would be practically vertical. 
This scarcity of normal matter in an NS ($I-I_v \sim10^{-3}I$) is not predicted by realistic models of NS structure. Hence, we may assume that Vela was in phase 1 -- phase 3 is excluded since phase 2 is not a viable option. 
However, this also seems unlikely since this event immediately precedes the fast spin-up, where enough lag has to be positive in order to store momentum for the glitch: within the standard scenario, vortices can undergo inward creep (so that the mutual friction torque acts in the same direction as the external torque) only if the local lag is momentarily reversed.  

In conclusion, such a huge anti-glitch precursor challenges the standard cartoon of pulsar glitches, unless one invokes the occurrence of the \emph{inward creep} phase just before the main spin-up event. 
Since we only used the general equation \eqref{qser}, assuming complex models for the mutual friction or adding additional (fluid or rigid) components should not change this conclusion: in a region of the star there should be enough inward creep to have $\dt \Omega_p \ll -\oi$. 
\\
\\
\indent
\emph{\textbf{Possible explanations of the precursor} -} As proposed in \cite{erbil_vela_2020MNRAS}, the precursor could be a consequence of the formation of high vortex density regions surrounded by vortex depletion regions called \emph{traps} \cite{Cheng1988}. 
In this case, the displacement of vortices is a byproduct of a starquake, that acts as a trigger. According to this picture, the dislocations of the lattice induced by the quake provide stronger pinning sites for the vortex lines. These new sites can attract vortices over microscopic distances, hence the formation of traps. Since the velocity and direction of the displaced vortices depend on the concentration of dislocations, it is still not completely clear how the creation of traps can induce the average inward creep needed to justify the precursor. 

A tentative explanation in terms of stochastic fluctuations of the internal torque has also been proposed in \cite{ashton+2019} and \cite{carlin_brownian_2020}. If the continuous transfer of angular momentum from the superfluid to the normal component is particularly noisy, then the precursor may just be the effect of an abnormal fluctuation in the internal torque. Such a fluctuation increases in a short time the average lag between the components, pushing it above the unpinning threshold in some regions and triggering the subsequent glitch. Understanding if such internal torque fluctuations -- that, in turn, are fluctuations in the collective creep motion of many vortices -- can actually take place in a real pulsar is key to falsifying this possibility. Moreover, the statistical occurrence of such large fluctuations should be closely related to the observed statistical occurrence of glitches, or a subset of all the glitches observed in a pulsar.

%Future observations of well-resolved glitches have, thus, the potential to challenge the current understanding and, possibly, to inspire new physical  paradigms for glitch modelling. 
Another possible explanation of the precursor is based on the fact we do not directly observe $\Omega_p$ but rather a signal generated in the magnetosphere of the pulsar, that has its own dynamics. 
Hence, the precursor may not represent a real change in $\Omega_p$. 
On the contrary, it could reflect a change of the emission region of the Vela, a ``magnetospheric slip'' that should be modelled together with $\Omega_p$ to fit the data \cite{montolimagistrelli+2020}. Recent analysis also revealed a flickering magnetospheric activity of the Vela around the 2016 glitch epoch \cite{ashton2020arXiv}. 
If confirmed, this would provide a further clue that the magnetosphere is indeed a complex dynamical region; see also \cite{bansgrove2020}, who simulated the magnetospheric perturbations induced by slipping faults during a crustquake.

\section{Constraints on neutron star structure}
\label{chap:constraints}
We now revise how to extract physical information from glitch observations. 
As often occurs in astrophysics, extrapolation is an indirect process, whose result depends on some microscopic input parameters and on the ability to build models that reproduce the relevant piece of physics. In the following, we describe two \emph{stationary}\footnote{
    Namely, their implementation does not require any detailed modelling of the internal dynamics and depends only on the stationary structure of the rotating NS.
}
constraints on the pinning force and entrainment. More speculative ideas based on dynamical modelling are also briefly addressed in Sec.~\ref{sec:stronger}. 
 
\subsection{Stationary constraint from the maximum glitch amplitude}
\label{sec:max_glitch}
 
Observations of glitches of large frequency jump $\Delta \nu$ can be used to test theoretical estimates of the pinning strength. Within the assumption of circular motion described in Sec. \ref{circo}, from \eqref{qser} we have that (ignoring the external torque that acts on much longer timescales)
 \begin{equation}
     2 \pi\Delta\nu = - \frac{I_v}{I} \langle \Delta\Omega_{vp}\rangle 
 \end{equation}
 where $\Delta\Omega_{vp}$ is the difference between the lag after and before the glitch, locally.
 The exact difference $ \Delta\Omega_{vp}$ is impossible to observe and depends on the history of the pulsar. 
 However, we can give a theoretical upper limit $\Omega^{*}_{vp}$ to the local lag $\Omega_{vp}$: defining as $f_P$ the maximum Magnus force per unit vortex length that pinning can sustain, from $|\mathbf{f}_M| =f_P$ with $\mathbf{v}_L = \mathbf{v}_p$ we obtain 
 \begin{equation}
      \Omega_{vp}(t,x,z)  <  \Omega^{*}_{vp}(x,z) 
      = \frac{f_P(r) (1-\epsilon_n(r))}{\kappa \, \rho_n(r) \, x} \, . 
 \end{equation}
At this point, an upper limit $\Delta \Omega_{\rm max}$ on the glitch amplitude can be obtained by assuming that the angular momentum reservoir was full before the glitch (its maximum capacitance being defined by the pinning strength), while it is fully emptied after the glitch:
 \begin{equation}
     2 \pi\Delta\nu < \Delta \Omega_{\rm max} =\frac{I_v}{I} \langle \Omega^{*}_{vp}\rangle  
 \end{equation}
We can write the above equation as (we recall that $\Delta\nu$ is the observed frequency jump)
\begin{equation}
    2 \pi\Delta\nu < \Delta \Omega_{\rm max} = \frac{\pi^2}{ I  \kappa} \int dr \, r^3  f_P(r)     \, .
    \label{risultato_omegaMax}
\end{equation}
This constraint depends neither on the entrainment parameter $\epsilon_n$ nor the density $\rho_n$, but it only depends on the radial profile $f_P(r)$ and, to a lesser extent, on the total baryon density $\rho(r)$ that defines the moment of inertia $I$. 
Clearly, the region where $f_P(r)>0$, namely the region that contributes to the integral in  \eqref{risultato_omegaMax}, should be contained within the region defined by $\rho^*_n(r)>0$. 
The constraint \eqref{risultato_omegaMax} has been derived in the limit of slack vortices, but the exact same result is valid also for the extreme case of straight vortices, see the discussion below~\eqref{omVeq}.

When general relativistic corrections  in the slow rotation approximation \cite{hartle67,andersson_comer01} are taken into account, the constraint \eqref{risultato_omegaMax} reads~\cite{antonelli_GR_2018}
\begin{equation}
    \Delta \Omega_{\rm max} = \frac{\pi^2}{I \kappa} \int_0^{R_d} \mathrm{d}r\, r^3\, \mathrm{e}^{\Lambda(r)}\, \frac{\mathcal{E}(r) + P(r)}{m_n n(r) c^2}f_P(r) \, ,
    \label{eq:max_glitch_GR}
\end{equation}
where $\Lambda$ is the metric function which describes length dilatation, $\mathcal{E}$ is the total mass-energy density and $P$ is the pressure and $n$ the total baryon density (the Equation of State, EOS, is $E=E(n)$ and $P=P(n)$). 
Here, the radial profiles are obtained by integrating the TOV equations, as the corrections to the stellar structure can be safely neglected for slowly rotating two-component NSs~\cite{andersson_comer01}. 

Again, the result in \eqref{eq:max_glitch_GR} depends only on the pinning force profile, the mass of the star and the EOS. Once the microphysical parameters like the EOS and the pinning forces are fixed, \eqref{risultato_omegaMax} constrains the mass of a pulsar. 
\begin{figure}
  \begin{minipage}[c]{0.65\textwidth}
    %\centering
    \includegraphics[width = 0.99\textwidth]{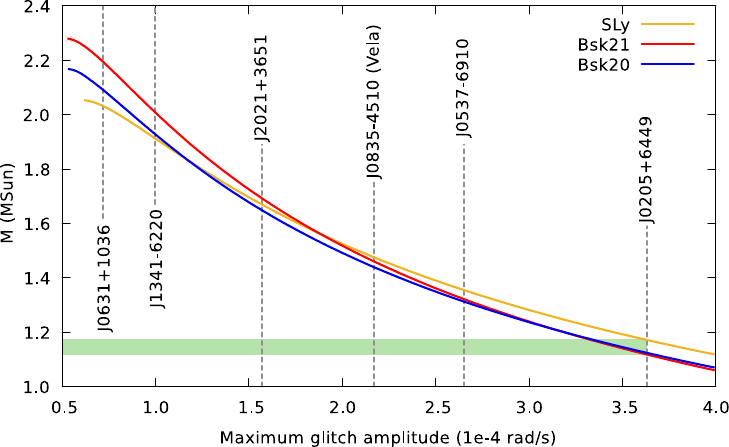}
          \end{minipage} \hspace{3mm}
      \begin{minipage}[c]{0.3\textwidth}
    \caption{The maximum glitch amplitude $\Delta \Omega_{\rm max}$ in equation \eqref{eq:max_glitch_GR} calculated as a function of mass for different unified EOSs \cite{antonelli_GR_2018}: SLy4 \cite{douchin_haensel01}, BSk20 and BSk21 \cite{goriely_etal10}. 
    The largest glitch displayed by some pulsars is also shown: the record-holder is J0205+6449 \cite{Livingstone2009ApJ}, which should have a mass lower than $1.1 - 1.2 \, M_\odot$, depending on the EOS employed.
    The pinning forces employed are the ones derived in \cite{seveso_etal16} for the inner crust.
    }
    \label{fig:max_glitch}
    \vspace{10mm}
     \end{minipage}
\end{figure}
This is represented in Fig.~\ref{fig:max_glitch}; not surprisingly, lighter pulsars can, in principle, undergo larger glitches. If we observe a large glitch of amplitude $\Delta \nu$ from a pulsar, all the inputs which predict a theoretical glitch smaller than the observed one are ruled out, at least for that object. To date, the pulsar that displayed the glitch of largest amplitude is J0205+6449 \cite{Livingstone2009ApJ}, that, therefore, should have a mass lower than $1.1 - 1.2 \, M_\odot$ when the pinning forces of \citet{seveso_etal16} are used in the inner crust. 
Since the theoretical minimum mass of a neutron star estimated from supernovae explosions is $\sim 1.17 \, M_\odot$ \cite{suwa_minimum_mass_2018}, this means that the input used to calculate $\Delta \Omega_{\rm max}$ in Fig.~\ref{fig:max_glitch} is only marginally acceptable. 
Hence, the largest glitch of J0205+6449 tells us that it may be necessary to extend the pinning region in the outer core, especially if even larger glitches will be detected in the future or new and more refined microscopic calculations will provide estimates of pinning forces in the crust that are lower than the ones used here. 

\subsection{Stationary constraint from the activity parameter}
\label{sec:activity}

Another type of information can be obtained by considering the so-called glitch activity, which represents the mean angular acceleration due to glitches \cite{McKenna1990Natur,lyne2000,espinoza2011,fuentes17}. 
For a pulsar that has undergone $N_{\rm gl}$ glitches of amplitude $\Delta\Omega_i$ at times $t_i$ with size $\Delta \Omega_i$ in a time span $T_{obs}$, we define the absolute activity as
\begin{equation}
\label{eq:aa}
\aa \, = \, \frac{1}{T_{obs}} \sum_{i=1}^{N_{gl}} \Delta \Omega_i \, .
\end{equation}
This definition works well under the assumption that the statistical properties of the stochastic sequence $(t_i, \Delta\Omega_i)$ do not depend on the window of observation (the sequence is wide-sense stationary), namely when the activities relative to sub-intervals are practically constant \cite{montoli_universe}. 
If $T_{obs}$ is not known and the only information available is the sequence $(t_i, \Delta\Omega_i)$, a quick way to estimate $\aa$  would be
\begin{equation}
\label{pizzasultana}
\aa  \, \approx \, \frac{1}{ t_{N_{\rm gl}} - t_1 }   
\left[ \sum_{i=2}^{N_{gl}-1}  \Delta \Omega_i + \frac{\Delta \Omega_1+\Delta \Omega_{N_{gl}}}{2} \right] 
\qquad \text{or} \qquad
\aa  \, \approx \, \frac{N_{\rm gl}-1}{ N_{\rm gl} (t_{N_{\rm gl}} - t_1) } \sum_{i=1}^{N_{gl}}  \Delta \Omega_i  
\, ,
\end{equation}
in order to correct the bias induced by the fact that the interval from $t_1$ to $t_{N_{\rm gl}}$ starts and ends with the detection of a glitch.
The second expression in \eqref{pizzasultana} is just a more democratic version of the first one, where the temporal ordering of the observed sequence is completely lost (the first and last glitch are not treated differently from the others), see App. A in \cite{montoli_universe}.

Due to small-number statistics, it is difficult to assess if an observed glitch sequence is wide-sense stationary. Luckily, there are a few obvious cases like the Vela and J0537-6910, where the cumulative glitch series has a clear and regular linear trend, see Fig. \ref{fig:fit_activity}. 
For instance, too few glitches have been detected in some pulsars, so it may not be safe to conclude that the inferred value of $\aa$ corresponds to the value that would be extrapolated if the observed sequence of glitches were longer\footnote{
    Furthermore, a large number of glitches $N_{gl}$ detected in a pulsar does not necessarily guarantee that the standard linear fit procedure provides a good estimate of  $\aa$~\cite{montoli+2020}. For example, consider a sequence where the first (or last) glitch has an amplitude much larger than the sum of all the others. A linear fit to the points of the cumulative amplitude will return a value of $\aa$ that is much smaller than the one simply calculated via \eqref{eq:aa}. This extreme example shows that the activities calculated in sub-intervals are not necessarily scattered around the value provided by~\eqref{eq:aa}, even for large $N_{gl}$.
}. 
In any case, the absolute activity is typically inferred by performing an ordinary linear regression on the cumulative glitch sequence, as in Fig.~\ref{fig:fit_activity}, even though this may lead to an underestimation of the associated uncertainty~\cite{mandel1957,Beck_fit_74,montoli_universe}.
\\
\\
\indent
\emph{\textbf{Constraint on the moment of inertia of the pinning region} -} 
\citet{link_1999} pointed out that the observed activity in a pulsar allows to constrain the moment of inertia fraction associated with the region that acts as an angular momentum reservoir. 
Let us revise this argument, that -- like the one for the maximum glitch size -- does not depend on the poorly-known details of the mutual friction, but only on the general statement \eqref{qser}.  
We formally divide $\Omega_p$ into the contributions due to the smooth and slow evolution observed during the waiting times  ($R$) and the fast one due to glitches ($G$), namely
\begin{equation}
\dot\Omega_p = \dot{\Omega}_{p}^R + \dot{\Omega}_{p}^G  \, .
\end{equation}
Now, we introduce an average over a long time interval of extension $T$,
\begin{equation}
[\![ \,  f \, ]\!]  \,  = \, T^{-1}\int_T  dt \, f(t,...)\,   
\label{average}
\end{equation}
so that, by definition, 
\begin{equation}
[\![ \, \dot{\Omega}_{p}^G \, ]\!] \, = \, \aa
\end{equation}
Consider now the total angular momentum balance \eqref{qser} and take its temporal average (assuming that changes in the moments of inertia are negligible):
\begin{equation}
[\![ I_v \langle \partial_t \Omega_{vp} \rangle \, ]\!] 
+ I \aa + I [\![  \,\dot{\Omega}_{p}^R \, ]\!] = -I \oi \, .
\label{puzzolo}
\end{equation}
The first term is always bounded due to the finiteness of the angular momentum reservoir,
\begin{equation}
[\![ I_v \langle \partial_t \Omega_{vp} \rangle \, ]\!] \,  = \,  
\frac{I_v}{T} \bigg(  \langle  \Omega_{vp} \rangle_{t=T} - \langle  \Omega_{vp} \rangle_{t=0} \bigg)
< \frac{I_v}{T}  \langle  \Omega^{*}_{vp} \rangle 
\, = \, \frac{I}{T} \Delta \Omega_{\rm max}
\end{equation}
where we have used the average in \eqref{average}. In this way, if $T$ is long enough to guarantee $\Delta \Omega_{\rm max}/T \ll \aa < \oi$, equation \eqref{puzzolo} can be approximated as
\begin{equation}
 \aa +  [\![  \, \dot{\Omega}_{p}^R \, ]\!] = - \oi \, .
 \label{piccionemalefico}
\end{equation}
We can use this expression to find a theoretical upper bound on the observed activity. To do so we need to set a robust limiting value to $\dot{\Omega}_{p}^R$. As discussed in Sec.~\ref{sec:dynamical_phases}, we can invoke the extreme scenario in which the vortices are pinned between glitches ($\dot{\Omega}_{v}^R=0$). 
This idealized situation is probably never realized in the real pulsar because of both thermal creep \cite{Alpar1984a} and disorder in the pinning landscape \cite{antonelli2020MNRAS}, so  we can safely conclude that
\begin{equation}
    [\![  \, \dot{\Omega}_{p}^R \, ]\!] \, < \, - \, \frac{I }{I-I_v} \, \oi \, . 
\end{equation}
Finally, plugging \eqref{piccionemalefico} in the inequality above gives the sought upper bound on the observed activity,
\begin{equation}
 \frac{\aa}{\oi} <  \frac{I_v}{I-I_v} \, . 
 \label{eq:constraint_activity}
\end{equation}
Formally, the above constraint reduces to the one derived by \citet{link_1999}  if we interpret $I_v$ as their $I_{res}$, the moment of inertia of the angular momentum reservoir (i.e., $I_{res}$ is the moment of inertia of the superfluid in the region where pinning is possible). For Vela, $\aa/\oi \approx 0.014$, indicating that the region inside the star that is associated with the momentum reservoir must carry at least $1.4\%$ of the total moment of
inertia, regardless of where this reservoir is.

The only difference between \eqref{eq:constraint_activity} and the result of \citet{link_1999} is that we used a formalism where entrainment is explicit. 
This entrainment correction was first derived by \citet{chamel_carter_2006MNRAS}: their result, albeit derived by using a slightly different -- but equivalent -- Newtonian formalism, is consistent with \eqref{eq:constraint_activity}. 
Minor differences are because we derived \eqref{eq:constraint_activity} without invoking the rigid-body assumption for the motion of the n-component.
However, the activity constraint is typically obtained
by assuming two rigid components with uniform angular velocities $\Omega_p$ and $\Omega_n$: by using this rigid-body framework and $I_v \ll I$, the constraint turns out to be \cite{chamel_carter_2006MNRAS,Andersson2012,Chamel2013,Delsate:2016}
\begin{equation}
 \frac{\aa}{\oi} <  \frac{I_n}{ 
 \big( 1-\langle \epsilon_n \rangle \big) \,  I }
 \qquad \quad
 I_n = \lim_{\epsilon_n \rightarrow 0} I_v 
  \, , 
 \label{eq:constraint_activity2}
\end{equation}
where $\langle \epsilon_n \rangle$ is the averaged entrainment in the reservoir region, see also \cite{SP10}. Therefore, we can consider the widely used constrain \eqref{eq:constraint_activity2} to be the rigid-body and small $I_n$ approximation of~\eqref{eq:constraint_activity}.

If we assume the pinned superfluid region to be limited in the crust of the star, and we fix the microphysical parameters, namely the EOS and the entrainment parameter, then the moments of inertia $I_v$ and $I$ in \eqref{eq:constraint_activity} depend only on the stellar mass. 
Therefore, similarly to the case for maximum glitch amplitude, here we obtain a constraint on the mass of the pulsar (see Fig.~\ref{fig:thecrust}). 
Apart from the EOS, both constraints depend on a single microphysical parameter: the maximum pinning force for the constraint from the largest glitch and the entrainment for the activity constraint.
Thus, the two different methods would allow for two tests on completely different microphysical parameters, if the real mass of the star were known.

\begin{figure}
  \begin{minipage}[c]{0.6\textwidth}
    %\centering
    \includegraphics[width = 0.99\textwidth]{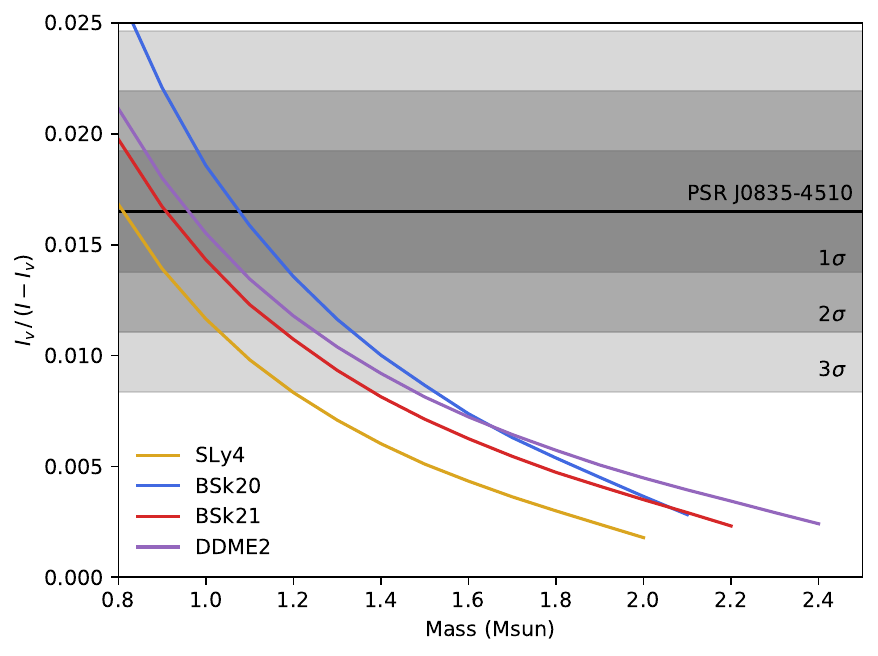}
          \end{minipage} \hspace{3mm}
      \begin{minipage}[c]{0.35\textwidth}
    \caption{
    The curve defining the activity constraint \eqref{eq:constraint_activity} for some EOSs. Here, the expression of $I_v$ is the general relativistic one in the slow rotation approximation, as defined in~\cite{antonelli_GR_2018,montoli_universe}. 
    The superfluid reservoir is limited to the inner and the entrainment is that of \citet{chamel2012}. The ratio $\aa/|\dot{\Omega}_\infty |$ is plotted for the Vela pulsar, along with the uncertainty as obtained via the statistical ``bootstrap'' procedure described in \cite{montoli_universe}.
    }
    \label{fig:thecrust}
    \vspace{10mm}
     \end{minipage}
\end{figure}

The entrainment parameter for the superflow in the crust of a neutron star has been obtained in \cite{chamel2012}, by calculating the effects of Bragg scattering due to the presence of the crustal lattice. The main result is a negative entrainment parameter $\epsilon_n \sim -10$ in the crust of the star, which implies a severely reduced superfluid density $\rho_n^*$ and a smaller $I_v$, making the requirement in \eqref{eq:constraint_activity} more difficult to be met. 
Studies on the activity parameter for the Vela yield an unrealistically low constraint on the mass of the pulsar for many EOSs, if a crustal reservoir and the entrainment calculated by \citet{chamel2012} are assumed \cite{Andersson2012, Chamel2013, Delsate:2016}. 
Several ways have been proposed to overcome this impasse, including the use of stiff EOSs \cite{pieka14}, a Bayesian analysis of the EOS uncertainty \cite{carreau+2019}, or an extension of the neutron superfluid participating in the glitch beyond the crust-core transition~\cite{Ho2015,montoli+2020}. 

Furthermore, the exact values of the entrainment parameter in the inner crust are still a matter of debate. Apart from calculations based on the effects of band structure on the effective neutron superfluid density in the crust  \cite{CC05,chamel2012}, other alternative estimates have been proposed, which include hydrodynamical approaches \cite{martinurban2016} or scattering in a disordered crustal lattice \cite{sauls+2020}, both of which yield milder entrainment effects in the crust. 
% The precise calculation of $\epsilon_n$ in the inner crust is particularly delicate \cite{martinurban2016,wantanabe17}, as its value can also be very dependent on the arrangement of the crustal lattice and the presence of disorder \cite{sauls+2020}.
Moreover, the effects of band structure -- that should lead to strong entrainment and a smaller value of $I_v$ -- are suppressed when the pairing gap is of order or greater than the strength of the crustal lattice potential, possibly reducing the entrainment in some layers of the inner crust \cite{wantanabe17}. However, the importance of this and other effects need further investigation, and a contained summary of the current issues regarding entrainment in the inner crust and core can be found in the review of \citet{chamel_entr_rev}.
In all these cases, the reduction of the neutron superfluid density would be less than estimated on the basis of band theory calculations. Therefore, it may be premature to rule out models of glitches where pinning occurs only in the crust.

Finally, different ways of estimating the activity parameter yield different values for the uncertainty of this parameter. 
For example, dropping the homoscedasticity hypothesis for the cumulative glitch amplitude data that define the glitch activity gives rise to larger uncertainties, see Fig.~\ref{fig:thecrust} and \cite{montoli_universe}.

\subsection{More stringent (but more model-dependent) constraints} 
\label{sec:stronger}

As soon as we tentatively extend the pinning region in the inner core, the moment of inertia $I_v$ grows so much that the mass of the Vela is consistent with any value smaller than the maximum one allowed by any reasonable EOS (the curves in Fig.~\ref{fig:thecrust} are considerably shifted upwards). 
However, observations of both the maximum glitch amplitude and the glitch activity may be used in tandem to put stronger -- but much less robust! -- constraints on the pulsar's mass. 
While the constraints coming from the maximum glitch amplitude and the activity are derived on the basis of the angular momentum balance only, we now have to model the internal dynamics, which amounts to selecting a particular mutual friction prescription, as tentatively done in~\cite{pizzochero17}. 

Starting from the corotation of the two components, forward integration in time of the dynamical equations \eqref{qser} and \eqref{omVeq} allows to follow the evolution of the  lag $\Omega_{vp}(x,z,t)$ and of the associated quantity $\Delta \Omega_m(t)$ defined as~\cite{montoli+2020}
\begin{equation}
    \Delta \Omega_m(t) \, = \, 
    \frac{I_v}{I} \langle \, \Omega_{vp} \, \rangle \, < \, \Delta \Omega_{\rm max} \, ,
    \label{galgano}
\end{equation}
that is nothing but a theoretical upper bound on the amplitude of a glitch triggered at time $t$ (assuming that at $t=0$ the star was in a state of corotation).  
In other words, $\Delta \Omega_m (t)$ sets a theoretical upper limit for the glitch amplitude at time $t$, given that a large glitch occurred at $t=0$. 
Since this is just an assumption to construct the theoretical upper bound $\Delta \Omega_m (t)$, it is not important if glitches in real pulsars typically empty the whole reservoir (probably not!).

The information coming from the glitch activity is used to roughly estimate the time $t$ needed to replenish the angular momentum reservoir to a level that is sufficient for the largest glitch amplitude $\Delta\Omega_{\rm obs}$ observed in the pulsar under study, $t_{act}\, \sim \, \Delta\Omega_{\rm obs} / \mathcal{A}_a$. Asking that $\Delta \Omega_m (t_{act})>\Delta\Omega_{\rm obs}$ provides the sough constraint on the pulsar's mass, once the microphysical input and the mutual friction prescriptions are fixed~\cite{montoli+2020}. Again, this is a test on the input used in the model rather than an estimate of a pulsar's mass.

Another, probably more promising, possibility is to adapt the activity constraint in \eqref{eq:constraint_activity} 
by taking into account the thermal history of a pulsar. 
\citet{Ho2015} used \eqref{eq:constraint_activity2} to constrain the mass of a sample of active pulsars by considering a superfluid reservoir that extends in the outer core, accounting for the temperature dependence of superfluidity. They inferred the temperature of pulsars from observations to constrain how much the superfluid reservoir extends into the core. 
This can be done by choosing a  model for the critical temperature of the neutron (singlet-state) superfluid as a function of baryon number density \cite{lombardo2001} and simulating the internal temperature of a pulsar. If the age of the pulsar is known -- or, at least, its characteristic age $P/(2 \dot{P})$ -- this gives an estimate of where the internal temperature of the cooling core is lower than the critical temperature for superfluidity. Hence, this approach makes it possible to obtain mass upper bounds by combining pulsar glitch data and the temperature dependence of superfluidity\footnote{
    The cooling and rotational evolution in an NS are coupled. For example,
    the late-time thermal evolution is affected by the spin-down due to the deposition of heat due to the action of mutual friction: besides the neutrino and photon emissions that are energy sinks, rotational energy is converted into heat by friction due to differential rotation of the superfluid \cite{page_1997ApJL,Larson_late_cooling_1999}. Conversely, the internal temperature also regulates the thermal creep \cite{Alpar1984a} and, possibly, localized heat deposition may increase the creep rate to initiate a glitch \cite{link1996,Larson2002MNRAS}.
}. 

Differently from \eqref{galgano}, this method does not assume anything about mutual friction, yet some extra information is needed. Instead of simulating the internal rotational dynamics, the difficulty is to simulate the cooling, which depends on several ingredients, most importantly the superfluid gaps \cite{pageCooling}. Moreover, the characteristic age is not always an accurate indication of the true age of a pulsar.
In the end, current mass estimates of glitching pulsars like the ones presented in \cite{Ho2015} and \cite{pizzochero17,montoli+2020} should be taken more as an indication of the influence of the assumed microscopic parameters on the final results. 

\section{Dynamical constraints from spin-up observations}
\label{sec:evolution}

We already discussed in Sec.~\ref{sec:2c} how observations of the slow relaxation after a glitch are the main evidence for the presence of superfluid components in NSs. Also, the short spin-up timescale associated with the 2000 and 2004 glitches in the Vela pulsar -- of the order of less than a minute \cite{dodson2002, Dodson:2007} -- and the particular slow-rise detected in the Crab pulsar \cite{shaw2018_largestCrab} can be interpreted as hints for the fact that the mutual friction is not simply linear, but that, on the contrary, its strength is modulated as the superfluid evolves across the various dynamical phases \cite{haskell2018MNRAS,erbil_crab_2019,celora2020MNRAS}. 
Despite this is the general picture at the basis of the vortex-mediated friction, the mutual friction is typically taken to be linear, e.g. \cite{haskell+2012,graber+2018,souriechamel2020,pizzochero+2020}. This may be a reasonable approximation, but only as long as the model is applied to a single dynamical phase.

The recent detection of a glitch in the Vela pulsar in December 2016  \cite{palfreyman+2018} has opened a window on the elusive spin-up phase. The peculiarity of this detection lies in the fact that this is the first pulse-to-pulse observation of a glitch in the act, allowing for unprecedented timing resolution of the spin-up and the following phases (see Sec.~\ref{sec:dynamical_phases}). 
Analysis of the phase residuals revealed that the initial fast spin-up occurs over a timescale $\tau_{\rm{r}} \lesssim 10 \, $s, as well as the presence of an \emph{overshoot} in the data in the first seconds of the glitch \cite{ashton+2019,pizzochero+2020}, a transient dynamical phase sometimes observed in glitch simulations that account for the non-uniformity of the superfluid velocity lag and stellar stratification \cite{Larson2002MNRAS,haskell+2012,antonelli17,graber+2018}. 

It is possible to use the minimal two-component model in \eqref{struzzo} to fit the data of \citet{palfreyman+2018} and extract a coupling timescale that, in turn, may be used to infer the mutual friction strength $\mathcal{B}_d$ during the spin-up. 
Taking the observed spin-up timescale to be $\tau_{\rm{r}} \sim 10 \, $s and assuming that the glitch is due to the superfluid in the crust -- the superfluid in the core is assumed to be corotating with the normal \cite{alpar84rapid}, so that $I_v/ I\sim 1\%$ -- from \eqref{tau_rigid} we conclude that\footnote{
    The estimate $\mathcal{B}_d\sim 10^{-5}$ given in equation (2) of \cite{ashton+2019} is based on the assumption of a single superfluid component that is limited to the inner crust ($I_v\sim 0.01 I$). Since they do not explicitly include entrainment, it should be reproduced by our $\langle \epsilon_n \rangle \ll 1$ case in~\eqref{onecomponent}. The two orders of magnitude discrepancy is due to a missing $I_v/I$ factor.
} 
\begin{equation}
\begin{split}
    \langle \mathcal{B}_d \rangle \sim 10^{-2} \qquad  & \text{if}  \quad \langle \epsilon_n \rangle \sim 10
    \\
    \langle \mathcal{B}_d \rangle \sim 10^{-3} \qquad  & \text{if} \quad \langle \epsilon_n \rangle \ll 1
\end{split}    
\label{onecomponent}
\end{equation}
in the inner crust, where we recall that $2 \Omega_p \sim 100 \,$rad/s for the Vela. 
The two cases refer, roughly, to the strong \cite{chamel2012,chamel_entr_rev} and weak \cite{martinurban2016,sauls+2020} entrainment cases, respectively.

The rough estimate in \eqref{onecomponent} depends on the fact that we reduced a fluid model to a much simpler simple model with only two rigid components. 
Clearly, such a simplification cannot be valid if, for example, the superfluid in the core is coupled to the normal component on a timescale that is comparable or longer to the one of the superfluid in the crust. Moreover, it is not possible to reproduce the overshoot feature since the general response \eqref{struzzo} of the two-component system can only fit an exponential rise\footnote{
    \citet{BAYM1969} introduced the model in Sec. \ref{sec:2c} to describe the post-glitch exponential relaxation, but the same model can be used to fit an exponential spin-up.
}.
 
\subsection{Models with three rigid components}

The minimal glitch model which allows for an overshoot requires at least three rigid components. This can be formally obtained from the two-component fluid equations \eqref{qser} and \eqref{omVeq} by splitting the whole superfluid domain into two non-overlapping regions \cite{montolimagistrelli+2020}. Taking the average over these two regions -- see the discussion below \eqref{tau_rigid} -- gives us the three-component analogue of the body-averaged model in Sec.~\ref{sec:2c},
\begin{align}
    	\begin{split}
		& x_p \dot{\Omega}_p + x_1 \dot{\Omega}_1 + x_2 \dot{\Omega}_2 \, = \, - |\dot{\Omega}_\infty|  \\
		& \dot{\Omega}_1 = - b_1 \left(\Omega_1 - \Omega_p \right)\\
		& \dot{\Omega}_2 = - b_2 \left(\Omega_2 - \Omega_p \right)
		\end{split}
		\label{eq:3c}
\end{align}
where $p$ indicates the normal (observable) component, $i = 1,2$ labels the two superfluid regions, $x_i = I_i/I$ is the moment of inertia fraction of the region $i$ and $b_i$ is the inverse of the coupling timescale between the $i$ region and the normal component. The normal fraction $x_p=1-x_1-x_2$ is not an independent parameter of the model.
The system \eqref{eq:3c} can be solved analytically. Following the same notation used in \eqref{struzzo}, the angular velocity residue  with respect to the steady-state spin-down law reads
\begin{equation}
    \Delta \Omega_p(t)  
    \, = \, 
    \Delta\Omega_p^\infty \left[ 1- \omega \, e^{-t \lambda_+} - (1-\omega)\,  e^{-t \lambda_-}  \right] \, ,
\label{eq:dOmegap}
\end{equation}
where $\omega$, $\Delta \Omega_p^\infty$ and $\lambda_\pm$ are functions of the parameters $x_{1,2}$, $b_{1,2}$ and $\oi$, as well as of the initial conditions for the two superfluids and the normal components. 
In particular, $\Delta \Omega_p^\infty$ is the angular velocity residue of the star at large $t$, after both exponential transients have fully relaxed. 

To extract information on the parameters $x_{1,2}$ and $b_{1,2}$, the formula  \eqref{eq:dOmegap} must be supplemented with the exact dependence of its four parameters $\omega$,  $\Delta\Omega_p^\infty$ and $\lambda_{\pm}$ on the physical parameters $x_{1,2}$, $b_{1,2}$ and the initial conditions. For example, the two relaxation timescales $1/\lambda_\pm$  in \eqref{eq:dOmegap} are  functions of all the four parameters $b_{1,2}$ and $x_{1,2}$, 
\begin{equation}
    \lambda_{\pm} = \frac{1}{2} \bigg[ \beta_1(1 - x_2) + \beta_2(1 - x_1) 
\pm \sqrt{ [\beta_1(1 - x_2) + \beta_2(1 - x_1)]^2 - 4 \beta_1 \beta_2 x_p } \bigg] \, ,
\quad \beta_i = b_i/x_p\, .
\end{equation}
The explicit formulae for $\omega$ and  $\Delta\Omega_p^\infty$, as well as the link of the averaged physical parameters $b_{1,2}$ and $x_{1,2}$ with the local microscopic input (including entrainment), are given in \cite{montolimagistrelli+2020}. 
\\
\\
\indent
\emph{\textbf{Physical interpretation of the parameters} -} The system \eqref{eq:3c} is agnostic about the location of the superfluid components: the moment of inertia fractions  $x_{1,2}$ do not necessarily correspond to the superfluid in the crust and the superfluid in the core. For example, the two superfluid components may be both located in the core of the star \cite{souriechamel2020}. The hope is to fit \eqref{eq:dOmegap} from observations of a well-resolved spin-up, and then to infer the most likely values for $x_{1,2}$ and $b_{1,2}$, so to constrain the superfluid regions and the (averaged) friction parameters.

In the original model of \citet{BAYM1969} -- the one defined in \eqref{baym1} and \eqref{baym2} -- the entrainment coupling was not explicit, but hidden in the expressions of the phenomenological parameters. We can reduce the full fluid problem to the set of phenomenological ordinary differential equations in \eqref{eq:3c} by following the same reasoning that led to \eqref{tau_rigid}. Now, instead of introducing a single average over the whole superfluid region, we formally split it into two non-overlapping regions, or layers, labelled by $i=1,2$:
\begin{equation}
     I_i = \int_i dI_v 
     \qquad \quad 
     \langle \, f \, \rangle = I_i^{-1} \! \! \int_i dI_v \, f  \,  
    \qquad 
    \text{where}
    \quad
    dI_v = d^3x \, x^2 \, \rho^*_n \, .
     \label{crispino_12}
\end{equation}
In this way, the inverse coupling timescales and the moments of inertia fractions can be expressed as (compare with equation \ref{tau_rigid})
\begin{equation}
\begin{split}
        & x_i = I_i/I \qquad \qquad \, \, x_{p} = (I-I_1-I_2)/I  
    \\  & \Omega_1 = \langle \Omega_v \rangle_1   \qquad \qquad  \Omega_2 = \langle \Omega_v \rangle_2
    \\  & b_i \approx \, 2  \Omega_{p}   \left\langle \, \frac{\mathcal{B}_d  \,  }{1-\epsilon_n} \right\rangle_i
    \approx \, 2  \Omega_{p}  \frac{  \langle \, \mathcal{B}_d  \,\rangle_i  }{\langle \, 1-\epsilon_n \, \rangle_i} 
\end{split}
    \label{tau_rigid_12}
\end{equation}
Again, the presence of entrainment affects the interpretation of the macroscopic parameters $x_i$ and $b_i$ in terms of the local input. 

It is worth remarking that the idea behind the reduction to a  model with rigid components is that the correlations in \eqref{eq:correlations_weak} are negligible. Hence, it may be a good strategy to identify the two regions in such a way that the resulting parameters $b_{1,2}$ are as different as possible, in such a way that the superfluid responds as uniformly as possible way within each region. 
Hence, it is reasonable to imagine that the two superfluid components represent regions with different physical properties (like the superfluid in the crust and the one in the inner core, but we will see that this is not the only possibility). 

Finally, the identities in \eqref{tau_rigid_12} also tell us that the description of the dynamical phases in Sec.~\ref{sec:dynamical_phases} is still valid, since  
\begin{equation}
     I_v  \, \langle \partial_t \Omega_{v} \rangle \, = \, I(x_1 \dot{\Omega}_1 + x_2 \dot{\Omega}_1)
    \qquad \qquad I_v = I(x_1+x_2) \, .
\end{equation}
In fact, the averaging procedure used to reduce a fluid model to a model with two or more rigid components reproduces perfectly the total angular momentum balance \eqref{qser}.

\subsection{Overshoot condition}

The maximum value $\Delta \Omega_{\rm over}$ of the angular velocity jump during a glitch can be obtained by finding the maximum value that can be attained in~\eqref{eq:dOmegap}. There are two possible situations: the maximum of $\Delta \Omega_p(t)$  is reached asymptotically -- namely $\Delta \Omega_{\rm over} = \Delta\Omega_p^\infty $ at large times --  or there exists a finite time for which  $\Delta \Omega_{\rm over} > \Delta\Omega_p^\infty $. In the latter case, the solution $\Delta \Omega_p(t)$ describes an overshoot and this happens if $\omega>1$. 
However, the parameter $\omega$ is, in general, a rather complex function of the initial lags and the physical parameters, so this overshoot condition is not of simple physical interpretation. 

It is possible to simplify the problem by limiting our attention to a particular subset of solutions, namely those arising from a physically-motivated choice of the unknown initial conditions: following~\cite{pizzochero+2020}, one of the two superfluid components, say $i=1$, is assumed to be a ``passive'' one that does not change its creep rate -- meaning that $b_1$ is constant before and after the triggering of the glitch. Hence, it is reasonable to set its initial lag to the steady-state value, $\Omega_{1}-\Omega_p = \oi/b_1$ at $t=0$. 
On the other hand, the component $i=2$ can be assumed to be the ``active'' one, namely the component that undergoes a transition from a pinned to an unpinned state. Since this is the component that is pinned before the glitch, it is also the component that provides the angular momentum for the observed spin-up. 
In this simplified setting, the mathematical condition for an overshoot is $b_1<b_2$: the post-glitch timescale $1/b_2$ of the active component (that in the pre-glitch state was only loosely coupled to the rest of the star because of pinning) must be smaller than the coupling timescale $1/b_1$ of the passive component~\cite{pizzochero+2020}.

From the physical point of view, the overshoot occurs when the superfluid region that stores the angular momentum for the glitch (the active component) can transfer its angular momentum to the normal component faster than the typical timescale the rest of the superfluid reacts with. 
This idea is further clarified by considering the limit $b_1 \ll b_2$: the passive component maintains its state of motion, being practically decoupled, while the active superfluid imparts its momentum to the normal component only, spinning it up. At this point, the lag between the passive superfluid and the normal component is smaller than its positive steady-state value or even possibly reversed.
The post-overshoot relaxation, during which the lag $\Omega_1-\Omega_p$ will tend to approach again the steady-state value, is modulated by the slower response of the passive superfluid (the $i=1$ component). 
This kind of behaviour is also clear from the simulations of the Vela 2016 glitch performed in \cite{graber+2018}, where the passive and active components are assumed to be the superfluid in the core and the one in the inner crust, respectively.

\citet{souriechamel2020} applied the model in \citep{pizzochero+2020} to study the possible impact of the pinning to the flux tubes in the outer core, where the magnetic field is expected to be predominantly toroidal and pinning to flux tubes may be the most effective \cite{GA14}. In their setting, the active component is the superfluid in the outer core, while the passive one is the superfluid in both the inner core and crust. In this way, they have been able to translate the simplified overshoot condition $b_1<b_2$ into a statement on the number of flux tubes that are, on average, pinned to a vortex. 
When the number of pinned flux tubes exceeds a certain threshold, the parameter $b_2$ is suppressed -- pinning tends to suppress mutual friction -- and no overshoot is possible.

\subsection{Constraint on the moment of inertia of the passive component}
 
By using a particular, simplified, solution of the three-component system in \eqref{eq:3c} derived in \cite{pizzochero+2020}, \citet{souriechamel2020} proposed a formula to constrain the moment of inertia fraction $x_1$ of the passive component from the observation of an overshoot in the angular velocity residue. This result turns out to be quite robust, as it can be derived directly from the general solution of \eqref{eq:3c}, including explicit entrainment couplings~\cite{montolimagistrelli+2020}. 

We have seen that an overshoot is likely to occur when $b_1 < b_2$. By expanding the value of $\Delta \Omega_{\rm over}$ in the limit $b_1 \ll b_2$, it is possible to show that~\cite{souriechamel2020,montolimagistrelli+2020}  
\begin{equation}
   %1- \Delta \Omega_p^\infty / \Delta \Omega_{\rm over} \,  < \, x_1 \, < 1 \, .
  x_1  \,  > \,  1- \Delta \Omega_p^\infty / \Delta \Omega_{\rm over}  
  \qquad \quad \textrm{if} \quad b_1 \ll b_2
    \label{x1_chamel}
\end{equation}
Hence, a measurement of both the asymptotic amplitude $\Delta \Omega_p^\infty$ and of the maximum amplitude at the peak of the overshoot $\Delta \Omega_{\rm over}$  provides a lower bound to the superfluid fraction $x_1$. 
For example, observing $\Delta \Omega_{\rm over } \approx 1.1 \times \Delta \Omega_p^\infty$ would immediately imply $x_1 > 10\%$, ruling out the possibility that the passive component is the inner crust alone.

Both $\Delta \Omega_p^\infty$ and $\Delta \Omega_{\rm over}$ may be obtained by fitting the data of a spin-up observation: we try to test the constraint \eqref{x1_chamel} with the data of the Vela 2016 glitch in Fig.~\ref{fig:constraint_chamel}. 
\begin{figure}
  \begin{minipage}[c]{0.5\textwidth}
    %\centering
    \includegraphics[width=0.9\textwidth]{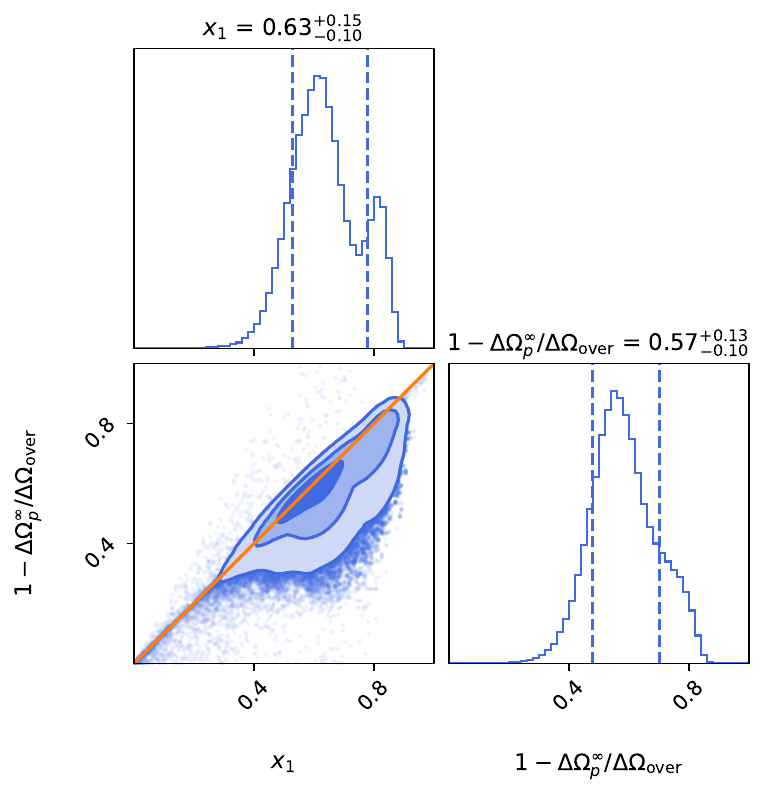}
          \end{minipage} \hspace{3mm}
      \begin{minipage}[c]{0.3\textwidth}
    \caption{ 
    Corner plot for the posterior distribution of the quantities $x_1$ -- the moment of inertia fraction of the passive superfluid component -- and $1-\Delta \Omega_p^\infty/\Delta \Omega_{\rm over}$. Both posteriors have been obtained by extending the Bayesian analysis of the Vela 2016 spin-up reported in \cite{montolimagistrelli+2020}.
    The orange diagonal represents the constraint proposed by \citet{souriechamel2020} and formally derived in \cite{montolimagistrelli+2020}: most of the posterior points satisfy the constraint
    imposed by equation \eqref{x1_chamel}, indicating the likely presence of a glitch overshoot in the data. 
    }
    \label{fig:constraint_chamel} \vspace{10mm}
     \end{minipage}
\end{figure}
The median of the probability distribution of $x_1$ is larger than that of $1 - \Delta \Omega_p^\infty / \Delta \Omega_{\rm over}$. In the covariance plot, most of the points fall in the region below the diagonal, thus respecting the constraint in \eqref{x1_chamel}. 
Since  \eqref{x1_chamel} is valid in the limit $b_1 \ll b_2$ but the analysis used to produce Fig.~\ref{fig:constraint_chamel} is valid for any value of $b_{1,2}$, this retrospectively confirms that the couplings  $b_{1,2}$ have values different by at least an order of magnitude. 
In fact, posteriors for $b_{1,2}$ reported by \citet{montolimagistrelli+2020} indicate that the
16th-84th percentile intervals are $b_1\sim 0.004 - 0.009\,$s$^{-1}$ and $b_2 \sim 0.08 - 24.64\,$s$^{-1}$.

\section{Conclusions}

%We reviewed some of the ideas behind the current understanding and modelling of pulsar glitches. 
Since the minimal glitch model of \citet{BAYM1969} and the seminal work of \citet{ANDERSON1975}, we now have a better understanding of how neutron star interiors can be described at the hydrodynamic level. 
While phenomenological models are useful to fit observations, information on the physics of the interior of an NS can be extracted only if we can provide a clear link between the fitted parameters and a more fundamental physical description of the system. 
Therefore, as a pedagogical exercise, we derived -- starting from a local fluid description that is more directly linked to the microphysics of the system -- the axially symmetric equations that constitute the backbone of almost all glitch models. In this way, the phenomenological parameters of the minimal model of \citet{BAYM1969}  -- or its natural multi-component generalization \cite{montolimagistrelli+2020} -- can be expressed in terms of the local properties of dense nuclear matter.

Timing observations of pulsar glitches can be used to test the microscopic input of glitch models. There are two robust (i.e., that do not depend on the poorly understood internal torques due to mutual friction) constraints coming, respectively, from the observation of glitches of large amplitude \cite{antonelli_GR_2018} and from the average glitch activity over an extended period of time \cite{link_1999,Chamel2013,andersson2013,Delsate:2016,pieka14,carreau+2019,montoli_universe}. 
Observation of large glitch amplitudes (the record holder is J0205+6449) can be used to test microscopic estimates of pinning forces, while the high activity of some pulsars (the Vela being the most active) provides a way to test the entrainment coupling. 
The two tests are completely independent -- apart from the fact that both rely on the choice of an EOS -- and seem to indicate that part of the superfluid in the outer core should participate in storing angular momentum: the upper limit on the inferred mass coming from the constraints is, for some objects, close or smaller than the theoretical minimum NS mass of $1.17\, M_\odot$ inferred from supernovae simulations \cite{suwa_minimum_mass_2018}.
This is interesting, as pinning of vortices to flux tubes could be a possible mechanism to store angular momentum~\citet{Alpar_pinningCore_2017}. 
The two stationary constraints may be refined but at the expense of solving the dissipative hydrodynamic equations (possibly coupled to heat) in the NS interiors. This has been done by invoking some simplified model that accounts for the mutual friction between the components~\cite{pizzochero17,montoli+2020} or by considering how the superfluid region evolves as the star cools \cite{Ho2015}, see the discussion in Sec.~\ref{sec:stronger}.

Information can also be extracted from a fit to the observed dynamics of a pulsar, see Sec. \ref{sec:evolution}.
In particular, the possibility of observing all the phases of a glitch in the act \cite{palfreyman+2018} justifies the efforts towards a better understanding of the processes that mediate the angular momentum exchange in a glitch.
For instance, our ability to extract physical information from glitch observations is directly linked to our ability to develop and solve reliable models for the internal torques between the components. This hard task boils down to understanding the vortex-mediated mutual friction between the superfluid components and the normal (observable) one during all phases of a glitch, as sketched in Fig.~\ref{fig:pinningladscape} and Fig.~\ref{fig:phases}.

One of the most important points is that the standard linear model of mutual friction \cite{mendell1991ApJ_II,epstein_baym92,andersson_MF}, is modified by the presence of pinning \cite{Alpar1984a,bulgac2013PhRvL,antonelli2020MNRAS}, quantum turbulence \cite{andersson_turbulence,Haskell_barenghi_2020}, long-range mutual interaction between vortices \cite{Howitt_Nbody_2020} and memory effects including hysteresis \cite{antonelli2020MNRAS,carlin2021ApJ}, namely kinetic processes that are out of equilibrium \cite{GAVASSINO_iordanskii}. 
All these issues regarding vortex-mediated momentum transfer still need further investigation and understanding. After all, aspects of creep and pinning are the object of active research even in laboratory superconductors, where the observed complex dynamics and memory phenomena arise from the balance between competing processes included in the equation of motion for the flux tubes~\cite{Blatter_review_1994,reich1999}.

As more resources will be devoted to the continuous monitoring of promising active pulsars -- like Crab \cite{shaw2018_largestCrab,basu_8glitches_2020}, the Vela \cite{palfreyman+2018} and J0537-6910 \cite{Ho_big_glitcher_2020} -- a whole new wealth of data will allow to test out theoretical understanding and, hopefully, to put interesting constraints on neutron star properties with the aid of next-generations glitch models. 
Planned multi-element telescopes -- such as the Square Kilometer Array \cite{stappers_SKA_2018} -- can also be extremely useful for high cadence pulsar monitoring, especially if multiple subarrays are available \cite{basu_8glitches_2020}.
When coupled with automated glitch detection pipelines, this could considerably increase the number and quality of detected glitches across the pulsar population, ushering in a new era of discovery in a field that has entered into its 50s~\cite{burnell_future}.

\section*{Acknowledgments}
Partial support comes from PHAROS, Cost Action CA16214, and INFN, the Italian Institute for Nuclear Physics. During the writing of the manuscript (before submission in August 2021), M.A. was supported by the Polish National Science Centre grant Sonata Bis 2015/18/E/ST9/00577 (B.~Haskell) and the Nicolaus Copernicus Astronomical Center,~Warsaw.
This revised version is dedicated to the memory of prof. Pierre Pizzochero (Milano - April 10, 2023).

%\onecolumn{
    %\bibliographystyle{apalike}
    \bibliography{Biblio}
  %  }

\label{lastpage}

\end{document}